%% file: ms.tex
\documentclass[12pt,preprint]{aastex}
\usepackage{graphicx}

\begin{document}

\title{Low Resolution Spectral Templates For Galaxies From 0.2 -- 10$\mu$m} 
\author{R.J.~Assef\altaffilmark{1},  
C.S.~Kochanek\altaffilmark{1},
M.~Brodwin\altaffilmark{2},
M.J.I.~Brown\altaffilmark{3},
N.~Caldwell\altaffilmark{4},
R.J.~Cool\altaffilmark{5},
P.~Eisenhardt\altaffilmark{2},
D.~Eisenstein\altaffilmark{5},
A.H.~Gonzalez\altaffilmark{6},
B.T.~Jannuzi\altaffilmark{7}
C.~Jones\altaffilmark{4},
E.~McKenzie\altaffilmark{8},
S.S.~Murray\altaffilmark{4},
D.~Stern\altaffilmark{2}
}

\affil{
  \altaffiltext{1} {Department of Astronomy, The Ohio State
  University, 140 W.\ 18th Ave., Columbus, OH 43210}
  \altaffiltext{2}{Jet Propulsion, California Institute of Technology,
  Mail Stop 169-506, Pasadena, CA91109}
  \altaffiltext{3}{School of Physics, Monash University, Clayton 3800,
  Victoria, Australia}
  \altaffiltext{4}{Harvard/Smithsonian Center for Astrophysics, 60
  Garden St., MS-67, Cambridge, MA 02138}
  \altaffiltext{5}{Steward Observatory, University of Arizona, 933 N
  Cherry Ave., Tucson, AZ 85121}
  \altaffiltext{6}{Department of Astronomy, University of Florida,
  Gainesville, FL 32611-2055}
  \altaffiltext{7}{KPNO/NOAO, 950 N. Cherry Ave., P.O. Box 26732,
  Tucson, AZ 85726}
  \altaffiltext{8}{Department of Physics and Astronomy, Colgate
  University, 13 Oak Drive, Hamilton, NY 13346}
}

\begin{abstract}
We built an optimal basis of low resolution templates for galaxies
over the wavelength range from 0.2 to 10 $\mu$m using a variant of the
algorithm presented by \citet{budava00}. We derived them using eleven
bands of photometry from the NDWFS, FLAMEX, zBo\"otes and IRAC Shallow
surveys for 16033 galaxies in the NDWFS Bo\"otes field with
spectroscopic redshifts measured by the AGN and Galaxy Evolution
Survey. We also developed algorithms to accurately determine
photometric redshifts, $K$ corrections and bolometric luminosities
using these templates. Our photometric redshifts have an accuracy of
$\sigma_z/(1+z)\ =\ 0.04$ when clipped to the best 95\%. We used these
templates to study the spectral type distribution in the field and to
estimate luminosity functions of galaxies as a function of redshift
and spectral type. In particular, we note that the 5-8$\mu$m color
distribution of galaxies is bimodal, much like the optical g--r
colors.
\end{abstract}


\section{Introduction}\label{sec:intro}

Imaging surveys are a very important and common tool in
astronomy. Large wide field surveys, such as the Two-Micron All Sky
Survey \citep[2MASS;][]{2mass} and the Sloan Digital Sky Survey
\citep[SDSS;][]{sdss}, and very deep ones, like GOODS \citep{goods}
and the Hubble Ultra Deep Field \citep{udf}, have radically improved
our understanding of the universe. The large galaxy samples yielded by
these surveys enable us, for example, to study the evolving space
density of galaxies \citep{bell04,brown07}, baryon oscillations
\citep{padma06} and the halo occupation distribution
\citep{zehavi05,ouchi05,lee06,white07,browninprep}. Astrophysical
applications of these surveys require measurements of quantities such
as the redshift, spectral type and rest frame and bolometric
magnitudes of the galaxies. Due to the enormous number or faintness of
the objects in these surveys, spectroscopic follow-up is extremely
expensive, if not impossible, for the great majority of the
sources. Even when spectra are available, they usually have low $S/N$,
so most estimates of these quantities still have to come from
broad-band photometry.

Extensive efforts over the last decade have shown that photometric
redshifts estimates from broad-band photometry are reasonably
accurate. Photometric redshift techniques can be divided into two main
families: methods based on empirical relations between color and
redshift that are usually implemented with neural networks
\citep[e.g.][]{wang98,brunner99,annz,connolly95}, and methods based on
Spectral Energy Distribution (SED) fitting techniques
\citep[e.g.][]{bolzonella00,benitez00}. The first family of methods
relies on the assumption that there is some relation between observed
properties of galaxies and redshift that can be empirically calibrated
using a training set of objects with both broad-band photometry and
spectroscopic redshifts. These methods can automatically accommodate
physical processes that are hard to model directly, such as dust
extinction and emission, but they cannot be used for estimating $K$
corrections, bolometric luminosities or redshifts outside the range of
the training set. SED fitting techniques rely on model spectra to
determine redshifts by minimizing the difference between observed and
expected broad-band colors. This family of methods does not have
redshift boundaries, as long as the observed rest-frame wavelengths
overlap those of the template SEDs, and they can be used to determine
$K$ corrections and bolometric luminosities. They typically have
larger uncertainties than the empirical methods
\citep[e.g.,][]{csabai03,brodwin06} and can fail badly for objects
poorly described by the templates.

Templates used by the SED fitting methods are either derived from
observations \citep[e.g.][]{cww80,kinney96} or from stellar population
synthesis models \citep[e.g.][]{bc93,bc03,pegase97}. Most of these
templates have limited wavelength coverage. In particular, the popular
\citet{cww80} and \citet{kinney96} templates do not extend into the
infrared and most synthetic templates have not been calibrated in this
range or lack physical processes that operate at these
wavelengths. Templates derived from observations sometimes come from
very noisy spectra \citep[e.g.][]{kinney96}, which could translate
small systematic errors into large errors in the broad band
colors. Templates from stellar population synthesis models do not
suffer from this problem, but sometimes do a poor job reproducing
observed properties of galaxies. For example, the red galaxy templates
of \citet{bc93} agree with observed optical colors, but severely
underestimate UV fluxes \citep[e.g., see Figure 4 of][]{donas95}, and
most models cannot reproduce the colors of star-forming galaxies
because they do not include or cannot model nebular emission, dust and
PAH emission features. While the \verb+Pegase.2+ models
\citep{pegase97} attempt to include these effects, their templates
have not been calibrated particularly far into the infrared.

\citet{budava00} and \citet{csabai00} developed a method that adjusts
template SEDs in order to overcome these problems. The method uses a
training data set to determine SEDs that accurately represent the
galaxies and then uses the updated SEDs for photometric redshifts, $K$
corrections and bolometric luminosities. A similar method has also
been developed by \citet[][also see \citealt{blanton06}]{blanton03},
focusing mostly on $K$ corrections, and by \citet{zebra06}, who
implemented it, along with other features, in their ZEBRA package.

In this paper, we derive low resolution spectral templates for
galaxies in the wavelength range 0.2--10 $\mu$m that accurately
reproduce galaxy SEDs. We derive them using the extensive photometric
observations of the NOAO Deep Wide-Field Survey
\citep[NDWFS;][]{ndwfs99} Bo\"otes field combined with the redshifts
from the spectroscopic observations of the AGN and Galaxy Evolution
Survey \citep[AGES;][]{ages} and a variant of the \citet{budava00}
method. AGES provides spectroscopic redshifts for approximately 17000
galaxies with $z\lesssim 1$, most of which have broad-band photometry
from 0.4 to 8 $\mu$m.

In \S~\ref{sec:data} we describe the data we use to obtain the
templates. In \S~\ref{sec:methods} we describe the method used to
derive the templates, as well as the algorithms used to determine
bolometric luminosities, $K$ corrections and photometric redshifts. In
\S~\ref{sec:results} we derive the templates and apply the algorithms
for $K$ corrections and photometric redshifts to the galaxies from the
AGES galaxy sample. And finally, in \S~\ref{sec:spec_clas}, we study
the spectral type distribution for approximately 65000 galaxies from
the NDWFS Bo\"otes field, based only on their photometry. We also use
photometric redshifts and $K$ corrections to determine luminosity
functions for this field. Throughout the paper we assume the
standard $\Lambda$CDM cosmology ($\Omega_{\rm M}=0.3$, $\Omega_{\rm
\Lambda}=0.7$, $\Omega_{\rm K}=0$ and $H_0=70$ km/s/Mpc).

\section{Data}\label{sec:data}

The NOAO Deep Wide-Field Survey is a deep optical and near-infrared
imaging survey that covers two 9.3 square degree fields, the Bo\"otes
and Cetus fields.  Both fields were imaged in $B_W$ (3500-4750 \AA,
peak at $\approx$ 4000 \AA), $R$ and $I$ pass-bands to depths
(5$\sigma$, 2$''$ diameter apt.) of approximately 26.5, 26, and 25.5
AB magnitude.  Both NDWFS fields have been completely imaged in the K
and K$_{\rm s}$ bands to a limiting AB magnitude of 21.

In this paper we focus on the Bo\"otes field observations, for which
there has also been extensive coverage at other
wavelengths. Specifically, we will also use the observations of the
Flamingos Extragalactic Survey \citep[FLAMEX;][]{flamex06}, which
covered about half of this field in the J and K$_{\rm s}$ bands, the
z' band observations of the zBo\"otes survey \citep{cool06}, and the
IRAC Shallow Survey \citep{irac04}, which observed the field with the
{\it{Spitzer Space Telescope}} Infrared Array Camera
\citep[IRAC;][]{fazio04} in Channels 1, 2, 3 and 4 (3.6, 4.5, 5.8 and
8 $\mu$m respectively). We will refer to this last four bands as C1,
C2, C3 and C4 respectively throughout the paper. It should be noted
that there are also radio (FIRST, \citealt{first};WENSS,
\citealt{wenss};WSRT, \citealt{devries02};NVSS, \citealt{nvss}),
far-IR (MIPS, \citealt{weedman06}), X-ray (XBo\"otes,
\citealt{murray05}) and UV ({\it{GALEX}}; \citealt{martin05})
observations of the NDWFS Bo\"otes field that we do not currently use.

The AGN and Galaxy Evolution Survey is a redshift survey in the NDWFS
Bo\"otes field. It has obtained spectra for $\approx$ 20000 objects in
the wavelength range from 3200\AA\ to 9200\AA\ with a resolution of
$R\approx 1000$ using the 6.5m MMT telescope and the 300 fiber robotic
Hectospec instrument \citep{fabricant05}. Spectroscopic redshifts have
been measured for about 17000 galaxies in the field with $0 < z <
1$. The median redshift is approximately 0.31.

We derive the templates using a total of 16033 galaxies with
spectroscopic redshifts and photometry in at least 6 of these 11 bands
[$B_W$, $R$, $I$ and K from NDWFS; z' from zBo\"otes; J and
K$_{\rm s}$ from FLAMEX; and C1, C2, C3 and C4 from the IRAC Shallow
Survey]. We required 6 bands so that we would always include some
combination of optical and IR photometry for each galaxy, but
requiring 5 or 7 would not affect our results. We use 6\farcs0
aperture magnitudes to derive the templates and SExtractor
\citep{sextractor96} Kron-like magnitudes for estimates of the total
flux. The photometry was corrected for Galactic extinction with the
\citet{schlegel98} model. We cannot easily distinguish between
non-detections and survey gaps from the existing photometry
compilations, so we make no use of upper bounds.

The magnitudes measured by NDWFS and FLAMEX are in the Vega
system. The IRAC magnitudes are in their own system, which is based on
the Kurucz model spectrum of Vega \citep[see][]{reach05}. The z'
magnitudes are in the AB system. Throughout the paper we keep these
conventions -- every magnitude computed is presented in its respective
system. We will refer to the objects with both photometry and
spectroscopic redshifts as the AGES galaxy sample.

\section{Methods}\label{sec:methods}

In this section we present the algorithms developed to build the low
resolution templates from the Bo\"otes field observations and estimate
$K$ corrections, bolometric magnitudes and photometric redshifts. We
have made the latter algorithms publicly available
\footnote{www.astronomy.ohio-state.edu/$\sim$rjassef/lrt} as part of a
Fortran-77 library that also incorporates other useful functions and
can carry out the calculations for any set of filters specified by the
user.

\subsection{Templates}\label{ssec:temp_proc}

We build our templates using a variant of the approach proposed by
\citet{budava00}. The flux $F_{i,b}$ of object $i$ in band $b$ is
given by
\begin{equation}\label{eq:flux}
F_{i,b}\ =\ c\ N_b\ \int_0^{\infty} \lambda^{-1}R_b(\lambda)\
f_i(\nu)\ d\lambda,
\end{equation}
\noindent where $N_b$ sets the normalization of the filter,
$R_b(\lambda)$ is the filter bandpass response per photon of
wavelength $\lambda$, $c$ is the speed of light, and $f_i(\nu)$ is the
object's observed spectrum measured in energy per unit area per unit
time per unit frequency.  In general, the spectra of a sample of
galaxies will not be fully independent of each other, but, instead,
can be regarded as different combinations of a small set, or basis, of
rest frame spectral templates $T_k (\nu)$. Thus, we can model the
observed flux of an object as
\begin{equation}\label{eq:mod_flux}
F_{i,b}^{mod}\ =\ c\ N_b\ \left(\frac{10\rm pc}{D_{l,i}}\right)^2\ \sum_k
a_{i,k}\ \int_0^{\infty}\ \lambda^{-1} R_b(\lambda)\ (1+z_i)\
T_k\left[(1+z_i)\nu\right]\ d\lambda,
\end{equation}
\noindent where $a_{i,k}$ is the contribution of spectral component
$k$ to the observed spectra, $z_i$ is the redshift of the galaxy and
$D_{l,i}$ is its luminosity distance. We have assigned a bolometric
luminosity of $10^{10} L_{\odot}$ and a distance of 10pc to the
template spectra (see \S~\ref{ssec:bol_lum_proc}). This relation can be
discretized as
\begin{equation}\label{eq:mod_flux_2}
F_{i,b}^{mod}\ =\ (1+z_i)\ c\ N_b\ \left(\frac{10\rm
pc}{D_{l,i}}\right)^2\ \sum_k a_{i,k}\ \sum_{\lambda_n} S_{i,b,\lambda_n}\
T_{k,\nu_n},
\end{equation}
\noindent where the $T_{k,\nu_n}$ are the discretized templates and 
\begin{equation}\label{eq:s_i_b_l}
S_{i,b,\lambda_n}\ =\ \int_{\lambda_n} ^{\lambda_{n+1}}\
\lambda^{-1} R_b[(1+z_i)\lambda]\ d\lambda
\end{equation}
\noindent is the sensitivity curve of filter $b$ shifted to the
redshift of the observed object and integrated over wavelength bin
$\lambda_n$.

The main idea of the method is to use the observed colors of galaxies
to fit for the spectral base components $T_{k,\nu_n}$ .
\citet{budava00} used as their initial guesses orthogonal spectral
components derived from a Principal Component Analysis (PCA)
decomposition of the \citet{cww80} galaxy templates (CWW from here
on). Keeping the best fit templates orthogonal to each other during
the iterative procedure, their final templates correspond to the
principal components of the observed galaxy spectra. One problem with
such a decomposition is that the model spectrum can be unphysical
(negative) in some regions unless there are priors on the permitted
values of the $a_{i,k}$.

\begin{figure}
  \begin{center}
    \plotone{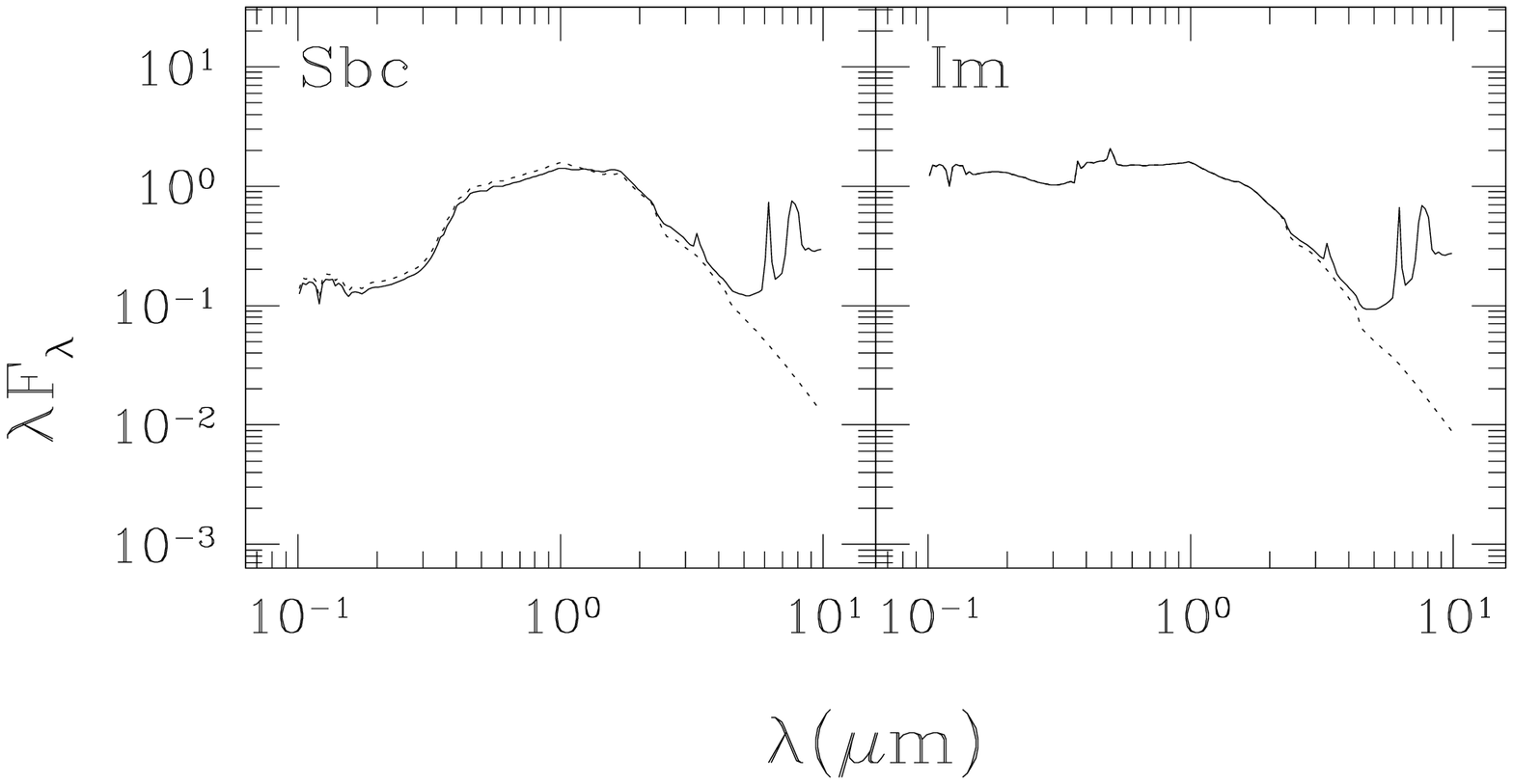}
    \caption{The solid lines show our initial guesses for the Sbc
    (left) and Im (right) templates, which were generated by extending
    the CWW templates to the mid-IR with the \citet{bc03} synthetic
    models and then adding the mid-IR part of the M82 and VCC 1003
    templates of \citet{devriendt99} to include dust/PAH emission
    features. For comparison, the dashed lines show the CWW templates
    extended into the mid-IR based only on the \citet{bc03} models.}
    \label{fg:cwwcomp}
  \end{center}
\end{figure}

Here we use an alternate approach that limits the construction of
unphysical spectra. We start from the Elliptical, Sbc and Im CWW
templates, extended to the mid infrared with the Bruzual and Charlot
synthetic models \citep{bc03}. To reproduce the mid-IR dust/PAH
features of star forming galaxies that these models lack, we spliced
onto the Sbc and Im models a combination of the mid-IR part of the
\citet{devriendt99} M82 and VCC 1003 templates, as shown in Figure
\ref{fg:cwwcomp}. We do not apply this modification to the Elliptical
template. Since all three templates represent very different star
formation histories (i.e. they have very different stellar
populations), they form a physical but not orthogonal basis set for
galaxy spectra. We will try to find the best modifications of these
spectra over the range 0.2 -- 10 $\mu$m which will fit the AGES
galaxies subject to the restrictions that the template spectra are
non-negative ($T_{k,\nu_n} \ge 0$) and that the spectrum of a galaxy
is a non-negative sum of these templates ($a_{i,k} \ge 0$). We will
refer to the templates as E, Sbc and Im throughout the paper since the
final optical spectra are sufficiently similar to the starting points
to retain the names.

Since we are building the template spectra with significantly higher
wavelength resolution than the broad band filters, we need to keep the
spectra from developing unphysical oscillatory structures during the
fit. We optimize the function
\begin{equation}\label{eq:G}
G\ = \chi^2\ +\ \frac{1}{\eta^2} H ,
\end{equation}
\noindent where the $\chi^2$ optimizes the fit to the templates, $H$
forces the templates to be smooth, and $\eta$ is a parameter that
determines the strength of the smoothing. The goodness of fit to the
data is
\begin{equation}\label{eq:chi2}
\chi^2\ =\ \sum_{i,b} \left(\frac{F_{i,b}\ -\
F^{mod}_{i,b}}{\sigma_{i,b}}\right)^2,
\end{equation}
\noindent where $F_{i,b}$ is the observed flux of object $i$ in band
$b$ with error $\sigma_{i,b}$, and the smoothing term 
\begin{equation}\label{eq:log_smooth}
H\ =\ \sum_{k,n} \left(\log
\frac{T_{k,\nu_n}}{Q_{k,\nu_n}} - \log
\frac{T_{k,\nu_{n+1}}}{Q_{k,\nu_{n+1}}}\right)^2 
\end{equation}
\noindent minimizes the logarithmic differences between the final
templates ($T_{k,\nu_n}$) and the initial templates
($Q_{k,\nu_n}$). If a too small value of $\eta$ is selected, the final
templates will not be very different from their initial guesses and
they will not be a good fit to the data. On the other hand, if a too
large value of $\eta$ is selected, the final templates will better fit
the data but they will show non-physical oscillatory
behavior. Selecting a value for $\eta$ between these two extremes
allows us to obtain galaxy templates that fit the photometry of the
sample better than the initial ones but are still well behaved. Since
the splices of the dust/PAH features are somewhat ad hoc, we decreased
the weight of the logarithmic smoothing linearly with wavelength from
1 to 10 $\mu$m.

Offsets in the photometry can potentially bias the final best fit
templates. Since our data covers a large range of redshifts, well
sampled in every filter, we can compute corrections to the nominal
photometric zero points of the AGES bands, as the overlapping regions
between filters should break any degeneracies. These adjustments
compensate both for the zero point errors and for any differences in
the effective photometric aperture created by the differing PSFs of
the observations. We can make these corrections to the extent that the
smoothing functions and the underlying templates we are trying to find
are not extremely different, since otherwise the smoothing can
compensate for the differences by introducing some large scale
behaviour into the zero point corrections rather than allowing the
templates to change. We assume that the zero point corrections are
small and not systematically related to each other, so all the large
scale behaviour in them should come from this degeneracy. We remove
any wavelength trend in the zero points by fitting a quadratic
function to the zero point corrections and then rescaling the
smoothing functions and the best fit templates.

We optimize equation (\ref{eq:G}) iteratively, starting with templates
matching the initial templates, $T_{k,\nu_n} = Q_{k,\nu_n}$. We then
iterate in steps: {\it{(a)}} estimate the galaxy weights $a_{i,k}$;
{\it{(b)}} estimate zero point corrections by adjusting $N_b$;
{\it{(c)}} sequentially optimize the templates and normalize them (see
\S~\ref{ssec:bol_lum_proc}); and {\it{(d)}} return to
{\it{(a)}}. After every five iterations, we remove the large scale
behaviour of the zero point corrections and rescale the smoothing
functions and templates. To optimize the templates we linearize the
smoothing term assuming that the change in $T_{k,\nu_n}$ is small
compared to $Q_{k,\nu_n}$. As the resulting equations are linear, we
can use a least squares algorithm in all the steps. Since we require
that every coefficient for which we fit is positive (all $a_{i,k}$,
$T_{k,\lambda}$ and $N_{b}$), we use the Non-Negative Least Squares
Solver (NNLS) of \citet{lawson74}. Our data sample contains objects
with bad data points or with heavy AGN contamination, so we adjust the
templates using only the 97\% of the galaxies with the best fits.

\subsection{Bolometric Luminosities and Template Normalization}\label{ssec:bol_lum_proc}

We normalize the templates to have a constant ``bolometric''
luminosity of $10^{10} L_{\odot}$ over the wavelength range from
$\lambda_{min} = 0.2 \mu$m to $\lambda_{max} = 10 \mu$m and to be at a
distance of 10pc. The ``bolometric'' luminosity we use is defined as
\begin{equation}\label{eq:lum_bol}
L_{bol}\ =\ 4 \pi D_l^2\ \int_{\lambda_{min}}^{\lambda_{max}} f(\nu)
\frac{d\lambda}{\lambda^2},
\end{equation}
\noindent where $f(\nu)$ is the observed SED of the object and $D_l$
is its luminosity distance. Since the normalizations of the templates
are the same, the total luminosity of a galaxy is simply
\begin{equation}\label{eq:lum_calc}
\frac{L_{bol}}{10^{10} L_{\odot}}\ =\ \sum_k\ a_{k},
\end{equation}
\noindent where the $a_k$ are the galaxy weight coefficients of
equation (\ref{eq:mod_flux}).

\subsection{$K$ Corrections}\label{ssec:kcorr_proc}

We can also use the templates to calculate $K$ corrections
\citep{oke68,hogg02} for virtually any band as long as it is inside
the wavelength range of the SED. This approach is similar to the one
taken by \citet[][also see \citealt{blanton06}] {blanton03}.

When observing a galaxy through a certain bandpass, the portion of the
rest frame SED of the object sampled by the bandpass will depend on
the redshift of the object. The $K$ correction can be defined as the
correction needed to transform the observed magnitude through bandpass
$b$ of an object at redshift $z$ to the magnitude we would measure
for an object with the same SED and the same apparent bolometric
magnitude but located at redshift $z_0$. We can write it as
\begin{equation}\label{kcorr_def}
m_b (z) = m_b (z_0) + K_b,
\end{equation}
\noindent with the $K$ correction $K_b$ defined as
\begin{equation}\label{eq:kcorr}
K_b\ =\ -2.5 \log{ \left[\frac{(1+z)}{(1+z_0)}\ \frac{\int_0^{\infty}
\frac{R_b (\lambda)}{\lambda} f[(1+z)\nu] d\lambda}{\int_0^{\infty}
\frac{R_b (\lambda)}{\lambda} f[(1+z_0)\nu] d\lambda} \right]},
\end{equation} 
\noindent where $f(\nu)$ is the rest frame SED of the object in units
of energy per unit area per unit time per unit wavelength. Usually,
$z_0 = 0$, so that the magnitude is corrected to the rest frame. One
alternative, adopted by the SDSS survey, is to set $z_0 = 0.1$,
corresponding to the mode of their redshift distribution, as this
minimizes the level of the corrections. Tables \ref{tab:3s_modmag} and
\ref{tab:4s_modmag} show the absolute magnitudes of the templates as a
function of redshift for the three and four templates model
respectively we discuss in \S~\ref{sec:results}. They can be used to
determine $K$ corrections for each of the AGES bands as well as other
commonly used ones (see captions for more information).

\subsection{Photometric Redshifts}\label{ssec:photoz_proc}

Once we have derived the templates, it is very easy to estimate
photometric redshifts for galaxies with fluxes $f_b$. For a given
redshift, we find the best combination of the basis templates by
minimizing
\begin{equation}\label{eq:chi2_photoz}
\chi^2(z,a_k)\ =\ \sum_b \left(\frac{f_b\ -\ c\ N_b\ (10{\rm pc}/D_l)^2\
\sum_k a_k (1+z) \sum_{\lambda} S_{b,\lambda}(z)
T_{k,\nu}}{\sigma_b}\right)^2 ,
\end{equation}
\noindent where $S_{b,\lambda}(z)$ is equal to $S_{i,b,\lambda}$ from
equation (\ref{eq:s_i_b_l}), to solve for $a_k(z)$. We continue to
require that $a_k(z) \geq 0$ and find the solution with the NNLS
algorithm. Then, with a grid search on the redshift values, we can
obtain the optimal redshift for the galaxy.

We included a luminosity prior in our model to avoid selecting
improbable luminosities as the best fits. Moreover, at very low
redshifts, luminosity is a better distance measure than color. We set
the probability for redshift $z$ to be
\begin{equation}\label{eq:p_z}
P(z)\ \propto\ e^{-\chi^2(z)/2}\ \Phi[M]\ dV_{com}(z),
\end{equation}
\noindent where $\Phi[M]$ is the luminosity function, the probability
per unit of co-moving volume for a galaxy to have absolute magnitude
$M$, and $dV_{com}$ is the co-moving volume per unit redshift as a
function of redshift. We assume the $R$-band luminosity function from
the Las Campanas Redshift Survey \citep{lcrslf}, which is parametrized
by a Schechter function \citep{schech76} with $\alpha = -0.7$ and
$M_{\star} = -21.4$. Our estimates might be improved by the use of
spectral type priors \citep{benitez00,zebra06}, but they are not
included in our present implementation. The $K$ corrections in Tables
\ref{tab:3s_modmag} and \ref{tab:4s_modmag} can also be used to
estimate photometric redshifts (see caption for more information).

\section{Results}\label{sec:results}

\subsection{Templates}\label{ssec:temp_res}

\begin{figure}
  \begin{center}
    \plotone{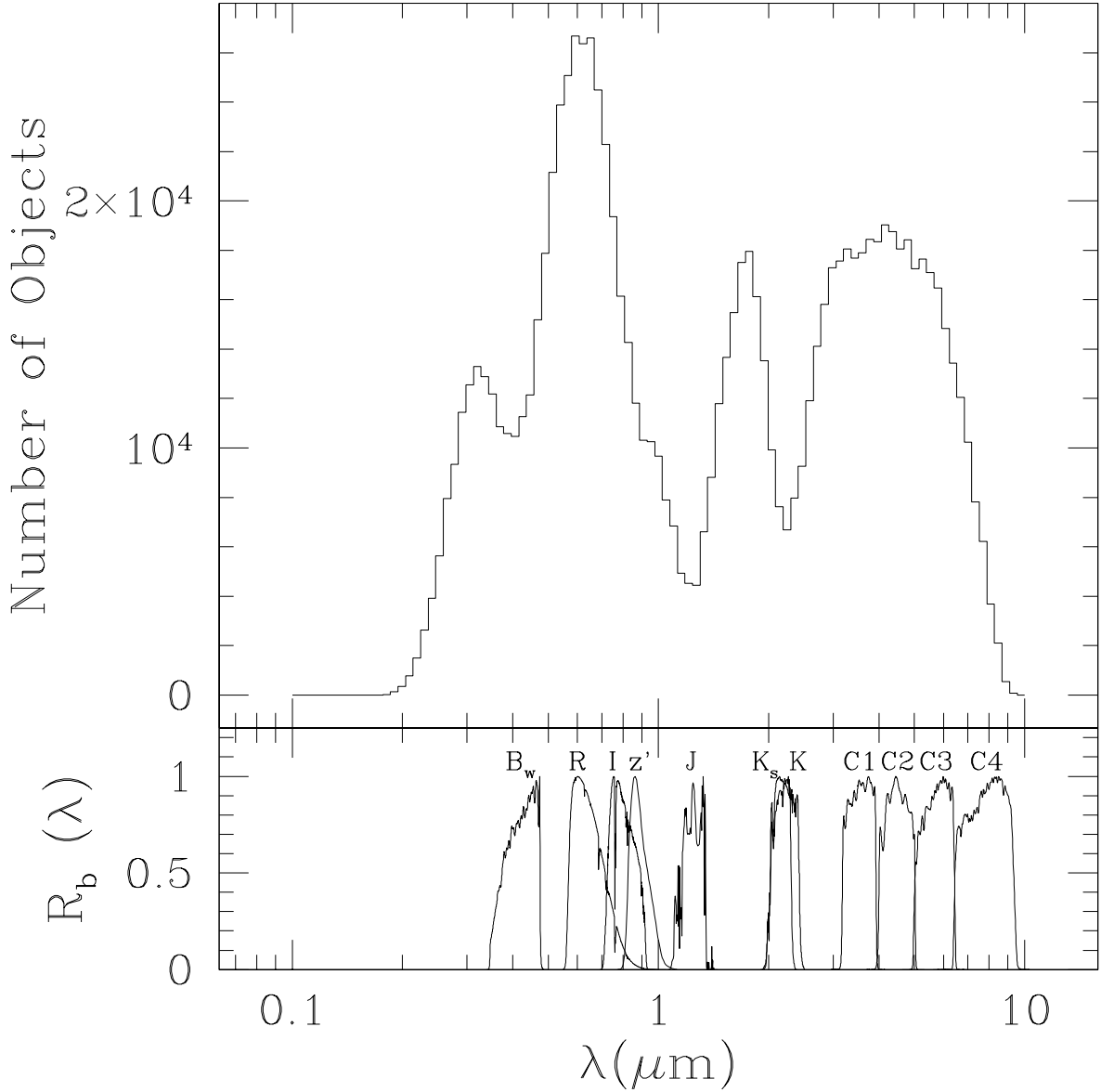}
    \caption{{\textit{(Top panel)}} The number of measurements used to
    derive the templates as a function of wavelength. We consider
    object $i$ to contribute to wavelength bin $n$ if
    $S_{i,b,\lambda_n}$ (as defined in eqn. [\ref{eq:s_i_b_l}]) is at
    least 10\% of its maximum in band $b$. {\textit{(Bottom panel)}}
    Filter sensitivity curves for the AGES bands. The dips seen in the
    top panel near 0.4, 1.2 and 2.1 $\mu$m are caused by the lack of V
    and H-band data and the significant gap between K and C1.}
    \label{fg:contrib}
  \end{center}
\end{figure}

Following the procedure outlined in \S~\ref{ssec:temp_proc}, we fit a
model based on the three modified CWW templates described in
\S~\ref{ssec:temp_proc} to the AGES galaxy sample, using the
photometry for the eleven bands described in \S~\ref{sec:data}. The
top panel of Figure \ref{fg:contrib} shows the number of objects used
to derive the templates as a function of wavelength and the response
curves of our eleven filters. The peaks in Figure \ref{fg:contrib}
correspond to the mean wavelengths of the filters displaced by the
redshift mode of our sample ($\sim 0.2$). Given our standard template
resolution, 160 logarithmically spaced wavelengths from 0.2--10
$\mu$m, these models have $N_{\rm DOF}=90669$ degrees of freedom. We
fit the data assuming the magnitude uncertainties are the larger of
the measured errors and 0.05 mag. This minimum error was imposed so
that low redshift galaxies with very small formal uncertainties did
not dominate the fits.

\begin{figure}
  \begin{center}
    \plottwo{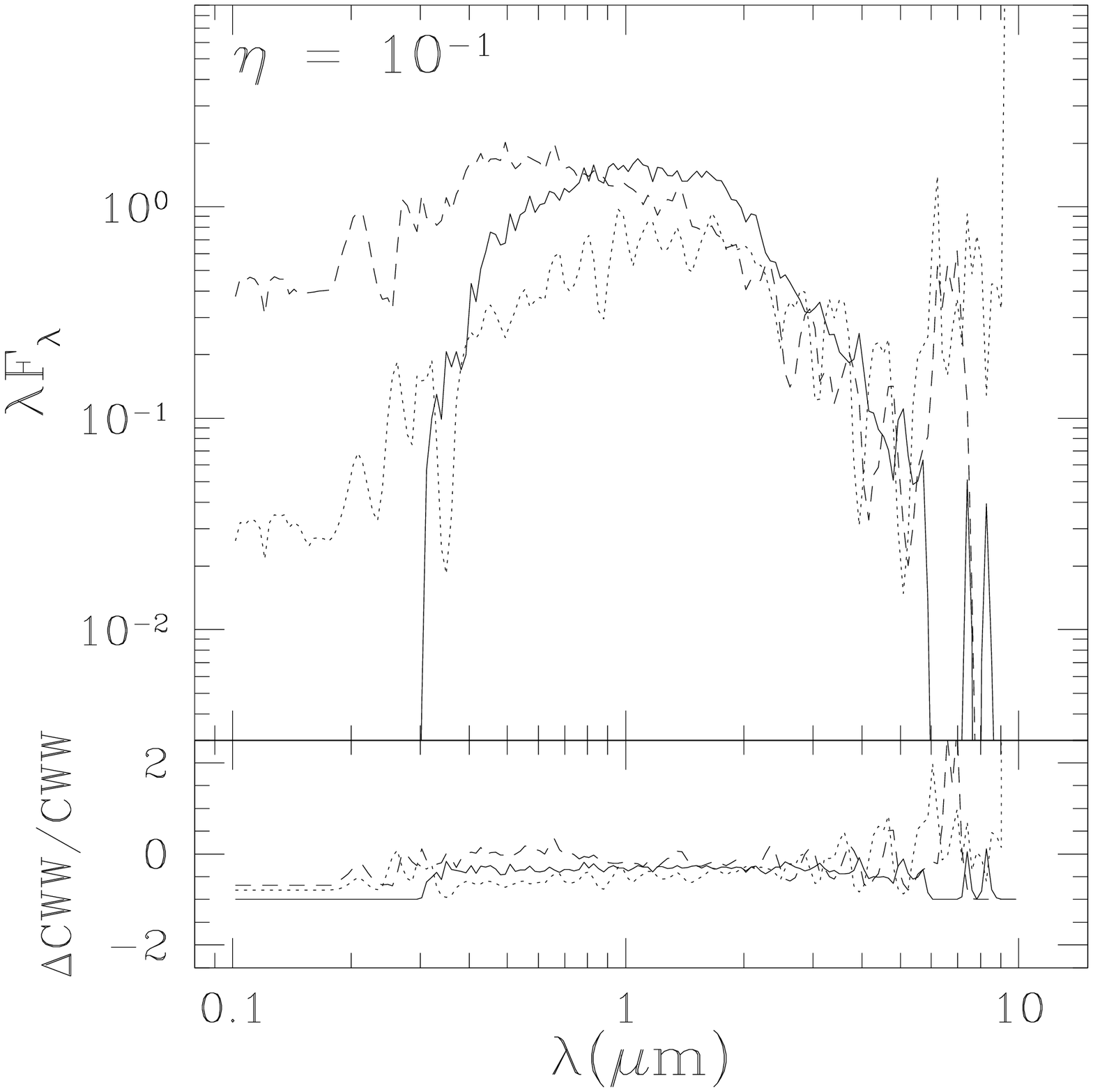}{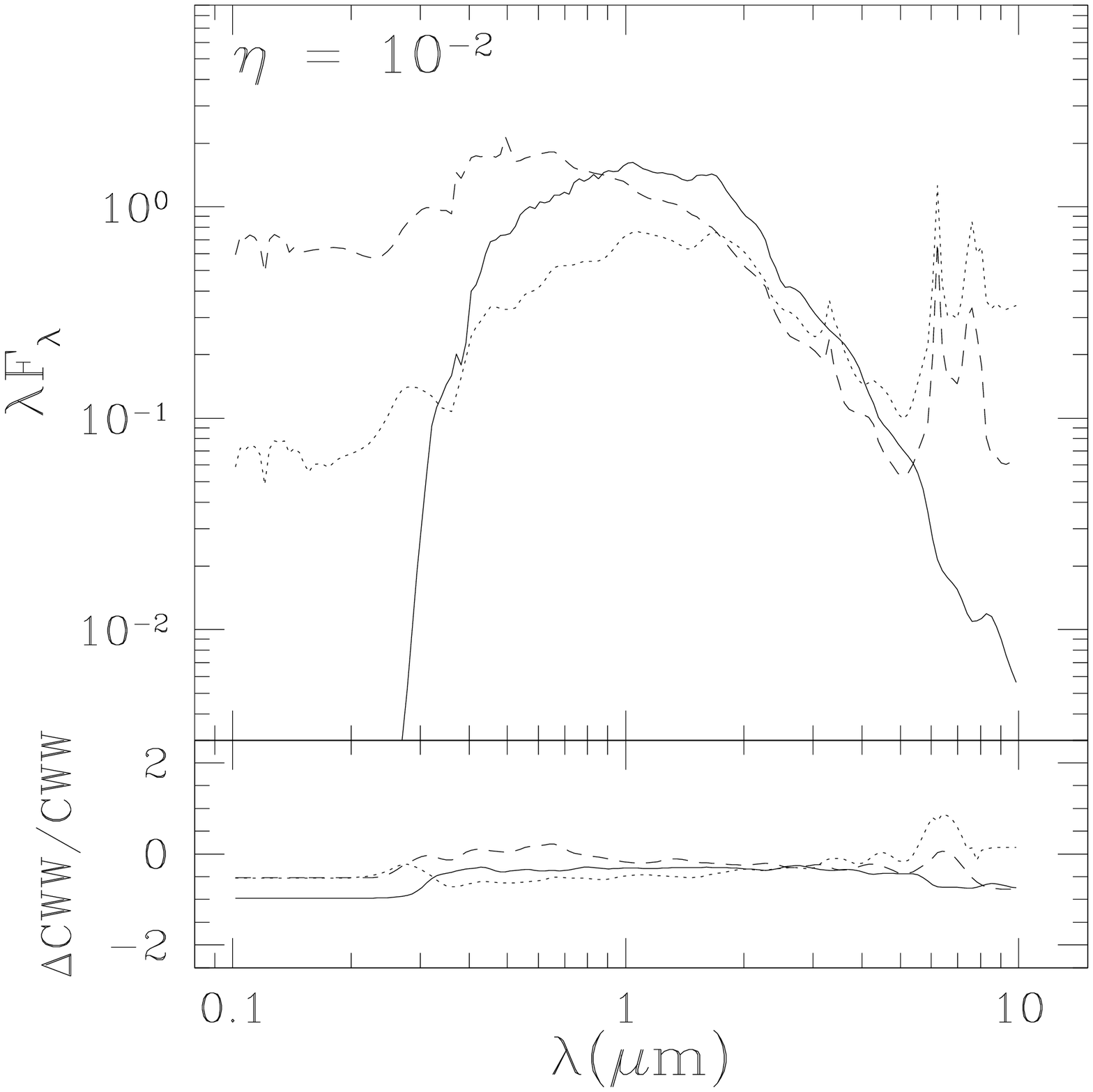}
    \plottwo{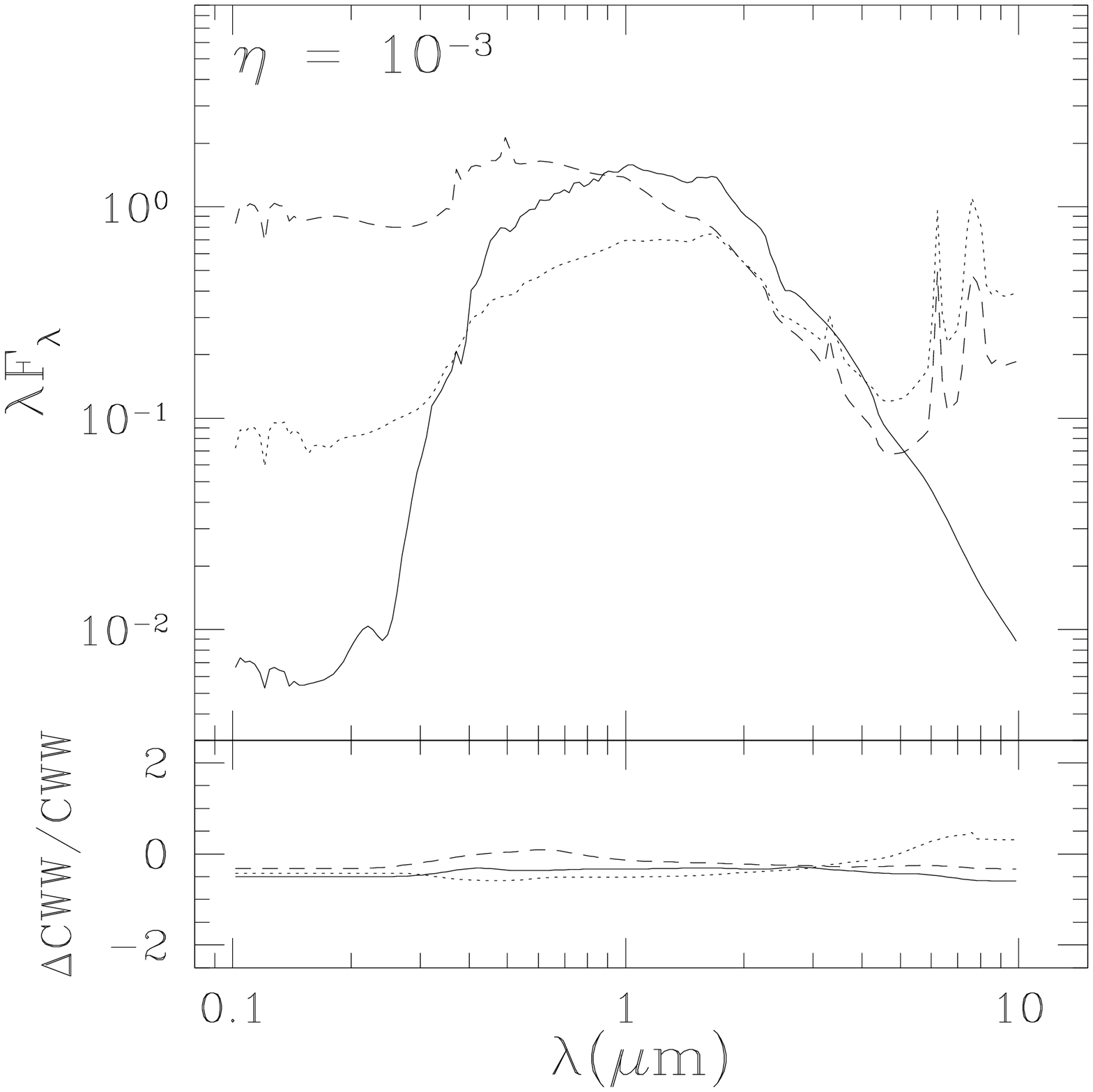}{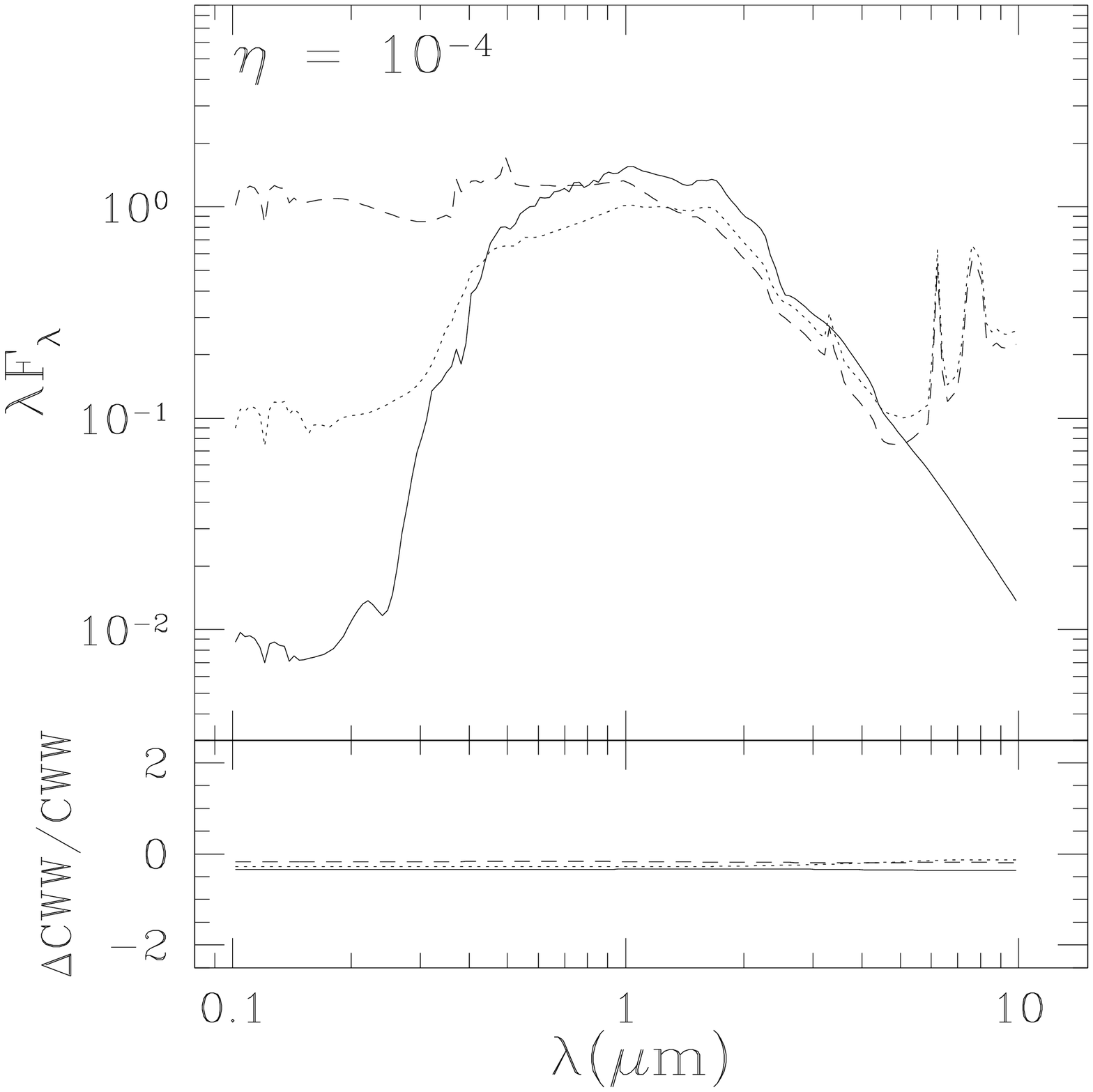}
    \caption{{\textit{(Top sub-panels)}} The Elliptical (solid line),
    Sbc (dotted) and Im (dashed) templates as a function of
    the smoothing strength weight $\eta$. Lower values of $\eta$
    correspond to stronger smoothing. {\textit{(Bottom sub-panels)}}
    The fractional change in the templates from the extended-CWW
    templates. For each case, we have approximately 90600 degrees of
    freedom.}
    \label{fg:allspeceta}
  \end{center}
\end{figure}

\begin{figure}
  \begin{center}
    \plotone{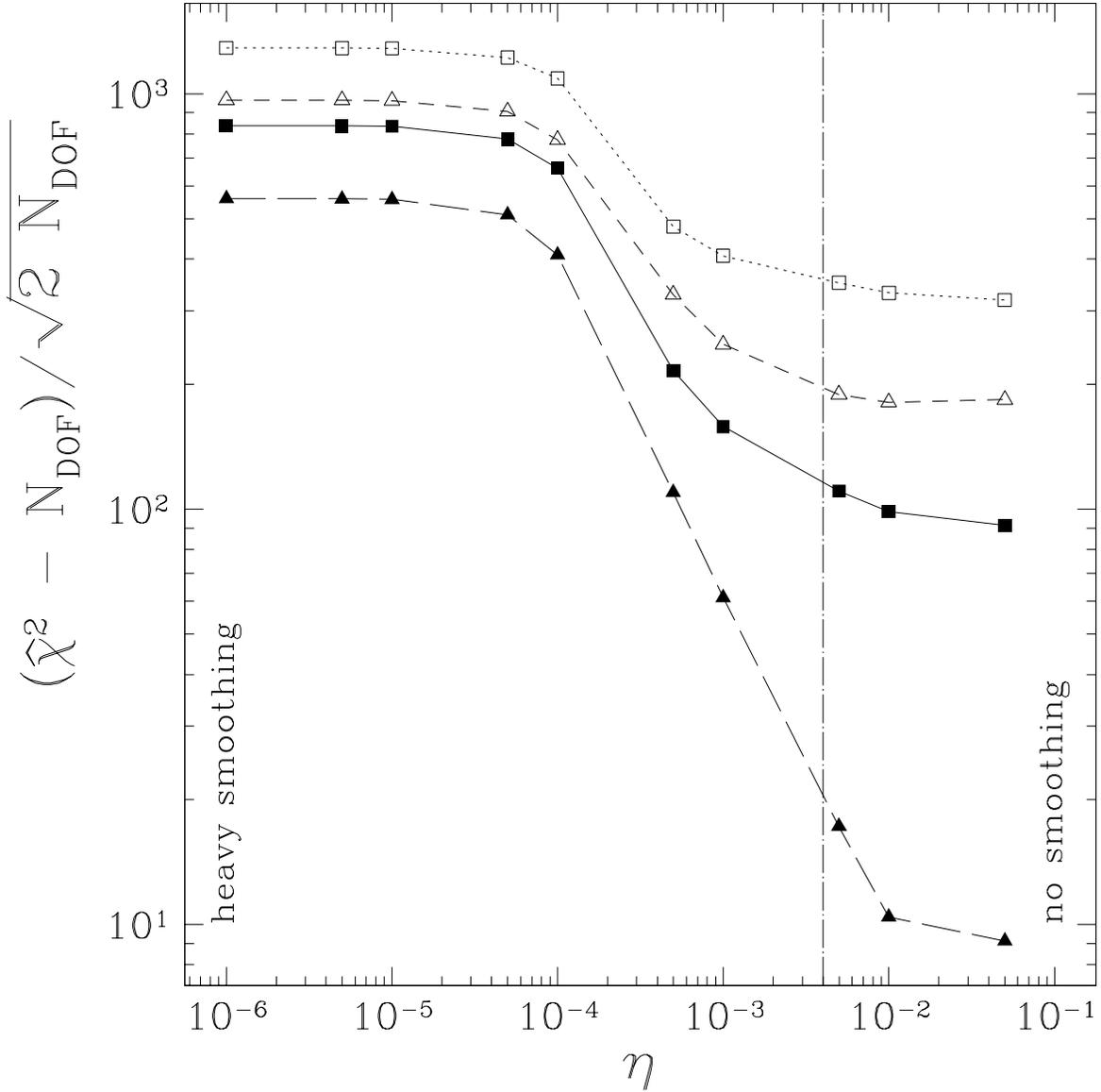}
    \caption{The deviation of the goodness of fit $\chi^2$ from the
    number of degrees of freedom $N_{\rm DOF}$ in units of the
    expected standard deviation in the absence of smoothing, $\sqrt{2
    \rm N_{\rm DOF}}$, as a function of the smoothing strength
    parameter $\eta$ for the three {\it{(squares)}} and four
    {\it{(triangle)}} template models. The selected value of
    $\eta=0.004$ is indicated by the vertical line. In the plot,
    $\hat{\chi}^2$ stands for the normalized $\chi^2$ such that if
    $\eta \to \infty$, $\hat{\chi}^2 = \rm{N_{\rm DOF}}$ for the four
    template model. The filled points show the results for the
    subsample used to build the templates, where we drop the 3\% of
    galaxies with the worst fit, and the open points show the results
    including all objects. That those 3\% of galaxies contribute more
    than 70\% of the total $\chi^2$ justifies their elimination.}
    \label{fg:chinorm}
  \end{center}
\end{figure}

\begin{figure}
  \begin{center}
    \plotone{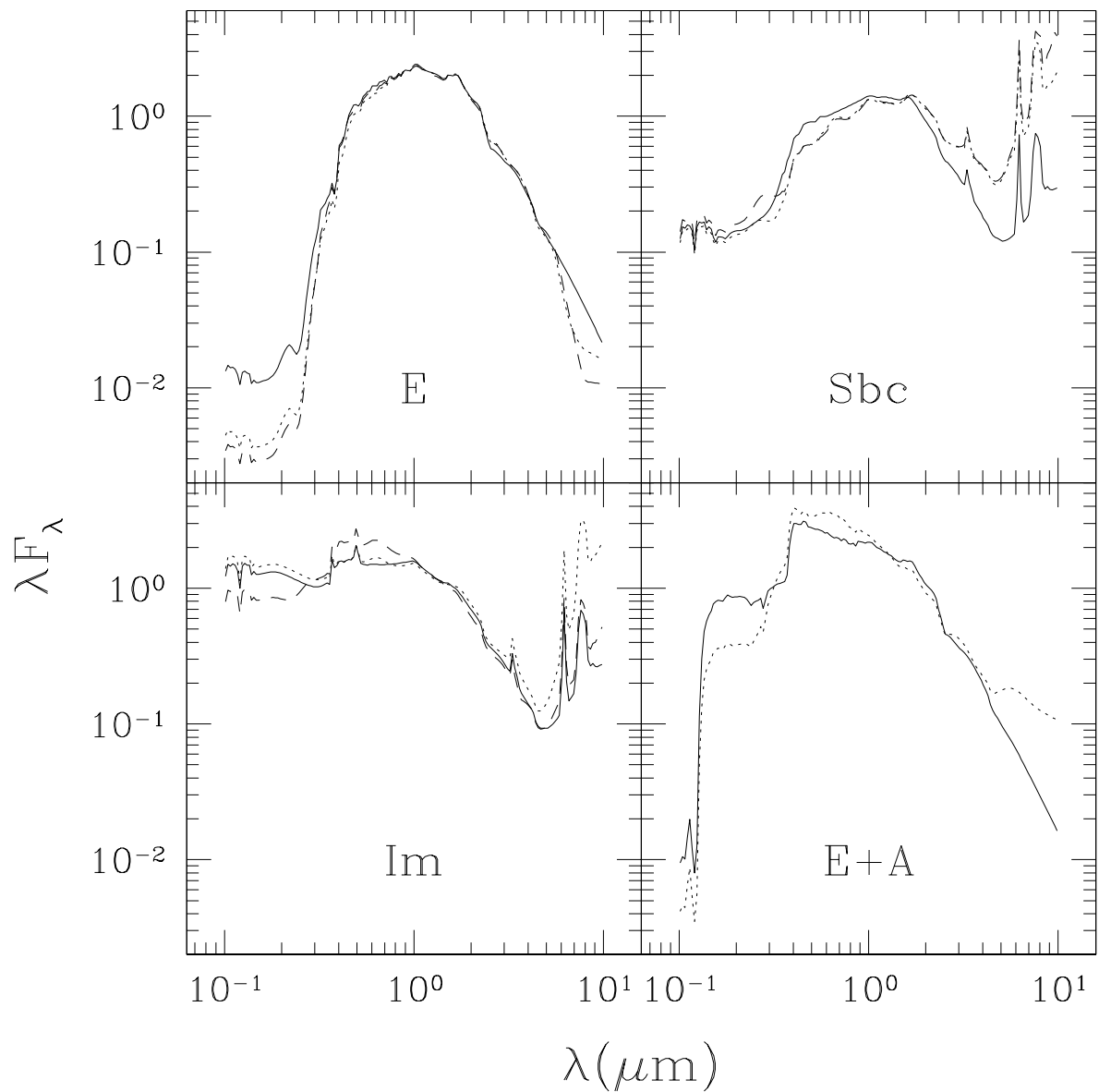}
    \caption{The templates derived using the algorithm described in
    \S~\ref{ssec:temp_proc} for the three {\it{(dashed)}} and four
    {\it{(dotted)}} template models compared to their initial guesses
    {\it{(solid)}}. All templates are normalized so that they have the
    same integrated energy from 0.5 to 2$\mu$m.}
    \label{fg:allspeccomp}
  \end{center}
\end{figure}

To choose an appropriate smoothing weight $\eta$, we first fit the
templates for a range of values. Figure \ref{fg:allspeceta} shows the
best fitting templates for different weights $\eta$, the $\chi^2$ of
each fit and the residuals when compared to their initial
guesses. In an ideal world, we would simply use the value of
$\eta$ that gives $\chi^2/N_{\rm DOF}=1$. Unfortunately, we have
imperfect errors for the data (e.g. bad data points and systematic
errors from seeing variations) and imperfect templates that cannot
encompass all physical parameters of real galaxies, so we are forced
to adjust $\eta$ on an empirical basis. Fortunately, the results are
not very sensitive to our choice provided it is reasonable. With
little smoothing ($\eta=0.1$) we obtain a relatively low $\chi^2$ but
find very unnatural, rapidly oscillating spectra. On the other hand,
very heavy smoothing ($\eta=10^{-4}$) gives spectra that are not
significantly different from their initial guesses and have
significantly higher $\chi^2$. Figure \ref{fg:chinorm} shows the
goodness of fit as a function of the smoothing weight, where we use a
renormalized fit statistic defined such that $\hat{\chi}^2\to N_{\rm
DOF}$ in the limit of no smoothing ($\eta\to \infty$). Clearly, we
want a value of $\eta$ near the zone of the steep decrease in
$\chi^2$. More specifically, we want a value of $\eta$ between
approximately $10^{-2}$ and $10^{-3}$ to ensure that the templates
have changed enough to fit the data well, but we have introduced no
unphysical oscillations. Since the photometric redshifts, the $K$
corrections and the bolometric luminosities are not very sensitive to
this parameter as long as it is on this range, we choose $\eta =
0.004$ for our standard models. The resulting templates are shown in
Figure \ref{fg:allspeccomp} and are provided in Table
\ref{tab:3spectab}. They produce a $\chi^2$ of 201414, which
for the 90669 degrees of freedom available gives $\chi^2/N_{\rm
DOF}=2.22$. The output templates are substantially different from our
initial modified CWW ones and wildly different from the \citet{bc03}
extended CWW templates. The fitted Elliptical template has a lower
ratio of optical and mid-infrared to near-infrared emission, and the
Sbc and Im templates have stronger PAH emissions in the mid-infrared.

Even though the three template model fits the data well, there is no
physical reason why three templates should be enough to reconstruct
the spectra for all galaxies in the sample. In particular, the initial
templates are either actually star forming (Sbc, Im) or have had no
recent star formation (Elliptical) -- there is no intermediate age
template. We tested a model with a 4$^{\rm th}$ template whose prior
was a CWW Elliptical template combined with an A0 stellar spectra from
the \verb+Pegase.2+ libraries \citep{pegase97} to mimic an E+A/K+A
spectrum. Since the dependence of the $\chi^2$ deviations should not
be extremely dependent on the types of templates that we are trying to
fit, we will use $\eta = 0.004$ as above. The resulting
templates are provided in Table \ref{tab:4spectab} and produce a
$\chi^2$ of 146410, which for the 75028 degrees of freedom available
gives $\chi^2/N_{\rm DOF} = 1.95$.

Figure \ref{fg:allspeccomp} shows the best fit three and four template
models compared to their initial guesses. They are clearly very
different from their initial guesses. While the best fit elliptical
and Sbc templates do not differ significantly from the previous case,
the Im is very different. Even though Figure \ref{fg:chinorm} shows
that adding an additional template significantly reduces the $\chi^2$
values, the formal improvement from adding the 4th template is only
about 19$\sigma$ based on the F-test. Moreover, as we shall see in
\S~\ref{ssec:photoz_res}, adding the extra component also creates
problems.

Compared to common template SEDs used in the literature, these
templates do a significantly better job of tracing the observed
color--color distribution of galaxies. Figures \ref{fg:color_diag_opt}
and \ref{fg:color_diag_mir} show the color distributions of the AGES
galaxies compared to the color ranges permitted by our basis of
templates in the optical and mid-IR bands respectively for four
redshift ranges. For comparison, we also show the optical color ranges
spanned by six commonly used templates: the CWW Elliptical, Sbc, Scd
and Im, and the \citet{kinney96} SB1 and SB2. The older templates
represent the colors of galaxies poorly, especially in the redshift
range 0.2--0.4, where the Sbc spiral template differs significantly
from the observations. Notice that they span lines instead of full
areas because they are single color points smeared by the redshift
range. This can be somewhat overcome by interpolating between the
templates, but this is highly dependent on the implementation of the
interpolation scheme. In the mid-IR, we show for comparison the colors
spanned by the \citet{bc03} extended CWW templates. These clearly do a
very poor job reproducing the observed color--color distribution. In
this same figure, note that the mid-IR distribution of galaxies at low
redshift is strongly bimodal, resembling the g--r color distribution
\citep[e.g.][]{strateva01,blanton03b,madgwick03,bell04}.

Finally, it should be noted that while fitting the templates we also
fitted for corrections to the nominal zero points of each of the AGES
bands, relative to the $B_W$ band. The zero points used are
3627.5, 3009.9, 2408.8, 3631.0, 1594.0, 666.7, 651.2, 277.5, 179.5,
116.6 and 63.1 Jy for the $B_W$, $R$, $I$, z', J, K$_{\rm s}$, K,
C1, C2, C3 and C4 bands respectively. The correction factors (relative
to $B_W$) are 1.00, 1.01, 1.02, 1.03, 0.97, 1.00, 1.00, 1.06,
0.98, 1.01 and 1.03 respectively (the large discrepancy for the IRAC
bands was also noted by \citet{brodwin06} and it seems to be related
to aperture corrections for the IRAC PSF). In general, these should be
viewed as corrections to a common mean photometric aperture rather
than errors in the zero-point calibrations. Note that we cannot
determine the absolute corrections since we are also fitting for the
fluxes of the galaxies. These corrections could be improved by
considering seeing variations between the individual observations, but
we will not pursue this question at present.

\subsection{$K$ Corrections}

As mentioned earlier, \citet{blanton03} followed an approach similar
to ours to determine $K$ corrections. To test our code, we compare our
$K$ corrections for the AGES galaxy sample with those from
the \verb+kcorrect v4_1_4+ code of \citet{blanton03}. Note that for
this comparison we use the 4 template basis model, as it provides a
better fit to the SEDs if the redshift is known (see
\S~\ref{ssec:temp_res} and \S~\ref{ssec:photoz_res}).

Figure \ref{fg:kcorr} shows the comparison for the $B_W$, $R$, $I$,
J, z and K bands at low ($z<0.3$) and high ($z>0.3$) redshift. We do
not examine the IRAC channels nor use them to fit the SEDs since
\verb+kcorrect v4_1_4+ cannot model mid-IR fluxes. In general, the
agreement is good, with a typical difference of less than about 0.1
magnitudes. The band with the largest dispersion is $B_W$. All
bands show some deviation in the mean of a few hundredths of a
magnitude, suggesting that there are some differences between the
templates used by the codes. Notice that there is a smaller deviation
at lower than at higher redshifts, which is expected since $K$
corrections tend to be bigger at higher redshifts and \verb+kcorrect+
was largely calibrated at lower redshifts than the AGES sample.

\subsection{Photometric Redshifts}\label{ssec:photoz_res}

Using the methods described in \S~\ref{ssec:photoz_proc}, we obtain
photometric redshifts for the AGES galaxy sample using the best fit
Elliptical, Sbc and Im templates described in \S~\ref{ssec:temp_res},
without considering the E+A component. We have so many sources that
there is no particular reason to have a separate training set. The top
left panel of Figure \ref{fg:zphot} shows a density contour plot of
the photometric redshifts, $z_p$, compared to the spectroscopic ones,
$z_s$, for the AGES galaxy sample. We show the dispersion in
$z_s$ at fixed $z_p$ since this is the distribution relevant for
characterizing the errors in photometric redshifts. The central
contours are tightly centered on the $z_p = z_s$ line, so the
algorithm works well for the typical galaxy. The results for this are
summarized in Table \ref{tab:zphot_sum}, as the ``3 templates/complete
sample'', where we give the standard dispersion
\begin{equation}\label{eq:sigmaz}
\left[\frac{\sigma_z}{(1+z)}\right]^2\ \equiv\ \frac{1}{N}\ \sum_{i=1}^N
\left(\frac{z_p^i - z_s^i}{1 + z_s^i}\right)^2 ,
\end{equation}
\noindent the median offset of $z_p - z_s$, the ranges of
$|z_p-z_s|/(1+z_s)$ encompassing 68.3, 95.5 and 99.7\% of the
distribution, and the dispersion $\Delta z$ defined by equation
(\ref{eq:sigmaz}) after clipping the sample to the 95\% of the
galaxies with the best $\left|z_p - z_s\right|$ to eliminate
outliers. The distribution of errors has very non-Gaussian tails. For
example, the region encompassing 68.3\% of the galaxies is 1.5 times
smaller than the dispersion. We explored the dependence of the
redshift errors on redshift, luminosity and color, finding that the
dominant effect is lower accuracy for bluer and fainter galaxies. For
example, if we sort the galaxies by their fitted SED elliptical
component fraction, $\hat{e}$, defined as
\begin{equation}\label{eq:ehat}
\hat{e} \equiv \frac{a_e}{a_e+a_s+a_i} ,
\end{equation}
\noindent where $a_e$, $a_s$ and $a_i$ are the Elliptical, Sbc and Im
template components of the galaxy SED, we find that $\sigma_z/(1+z) =
0.047$ for the galaxies with $\hat{e} > 2/3$ and $\sigma_z/(1+z) =
0.071$ for $\hat{e} < 1/3$.

Recently, \citet{brodwin06} estimated redshifts for galaxies and AGNs
in the IRAC Shallow survey using a hybrid algorithm between SED
fitting and empirical neural networks, calibrated with AGES
spectroscopic redshifts and photometry similar to that used in
here. Due to the lack of dust/PAH features in their templates, SED
fitting was only used for galaxies with $\rm C3 - \rm C4 < 1$, which
corresponds to galaxies with little or no star formation, while neural
networks were used for the rest of the galaxies and for the
AGNs. Eliminating the need to use different methods for star-forming
and quiescent galaxies was one of the motivations for our work. With
this hybrid approach, \citet{brodwin06} obtained $\sigma_z/(1+z) =
0.105$ and $\Delta z = 0.047$ for galaxies, about a factor of 1.8 and
1.2 larger than what we obtained, although the two galaxy samples are
not identical since \citet{brodwin06} used subset of AGES galaxies
with measured C2 magnitudes rather than the full sample.

We repeated the calculations using the 4 template model as shown in
the top right panel of Figure \ref{fg:zphot}. The distribution
statistics are again summarized in Table \ref{tab:zphot_sum}. The
dispersion when using four templates is equal to that for three
templates, while $\Delta z$ is larger. This seems to show that even
though the data are better fit using four templates rather than three,
the freedom introduced by including an extra template broadens the
photo-z distribution. Presumably this occurs because the four template
model allows colors that expand beyond the observed range for galaxies
(see Figs. \ref{fg:color_diag_opt} and \ref{fg:color_diag_mir}) while
the three template models do not.

We built the templates excluding the 3\% of galaxies most poorly fit
by them (see Figure \ref{fg:chinorm}). These poor fits are mostly
caused by extreme star formation, AGN contamination and bad
data. Figure \ref{fg:bad_ex} shows some examples of the worst and best
fit galaxies. The flat continuum in the mid-infrared is the signature
of an AGN \citep{stern05}. Figure \ref{fg:chi2z} shows that galaxies
that are poorly fit by the templates tend to have less reliable
photometric redshifts, so we examined the accuracy for galaxies whose
best fit photometric redshift yields a $\chi^2$ smaller than the 90th
percentile of its expected value. This criteria eliminates 25\% of the
original sample.  As summarized in Table \ref{tab:zphot_sum} and
illustrated in the bottom panels of Figure \ref{fg:zphot}, these
$\chi^2$-limited samples have distribution widths that are a factor of
1.2--1.4 smaller than for the sample as a whole, and by similar
amounts for the 68.3, 95.5 and 99.7\% intervals. We tried improving
the photometric redshifts for objects with bad data by sequentially
dropping individual magnitude measurements during the template
fitting. While this greatly improved the fits, the redshift accuracy
$\sigma_z/(1+z)$ worsened by 5--10\% when considering the full
sample. For objects that have AGN contamination, photometric redshifts
could be improved by adding an AGN template when the galaxy templates
fit poorly.

In these calculations we forced all $a_k$ coefficients to be positive
both while building the templates and while estimating the photometric
redshifts. It is possible that this limitation might worsen the
photometric redshifts, essentially by limiting the permitted range of
star formation rates. When we tested this by recalculating the
photometric redshifts without forcing $a_k \geq 0$, we found that the
dispersion in the redshifts increases by factors of 1.3 and 1.5 for
the 3 and 4 template models respectively. The problem is that the
added freedom allows the accessible color space to expand well beyond
that occupied by galaxies, thereby allowing good fits at bad
redshifts. We also investigated the effects of the luminosity priors
on the the photometric redshifts, and found out that while they
improve the accuracy, the gain is marginal (5--10\% effect in all
cases).

To further test our photometric redshift determinations, we obtained
the five bands of SDSS photometry for the galaxies in the AGES sample
and estimated their redshifts based solely on this information. We
find a dispersion of $\sigma_z/(1+z) = 0.086$ and $\Delta z = 0.052$,
again with highly non-Gaussian tails. While these values are worse
than what we had previously obtained for the same sample, as they are
based on a smaller number of photometric bands, they prove the
validity of our templates and algorithms. \citet{csabai03} estimated
photometric redshifts for galaxies brighter than $r'=18$ in the early
data release of SDSS with a method similar to ours and found a
standard deviation of $\sigma_{rms}=0.045$. If we limit our SDSS
sample to that magnitude, we find a very similar result of
$\sigma_{rms}=0.048$.

\section{Spectral Classifications}\label{sec:spec_clas}

We can now use the SED models for other applications, such as studying
the spectral distribution of the galaxies in the sample. For this
section, we use the full photometric galaxy sample of the Bo\"otes
field up to $I \leq 21$ mag, as derived from NDWFS, FLAMEX, zBo\"otes
and IRAC observations. This ``photometric'' galaxy sample consists on
approximately 80000 galaxies in total of which 69000 are usable
because they have photometry in at least four of the eleven bands
described in \S~\ref{sec:data} and have not been flagged by AGES for
either having an AGN contribution or being near a bright star. We
estimate photometric redshifts using the algorithm of
\S~\ref{ssec:photoz_proc}.

Figure \ref{fg:type_hist} shows the distribution of galaxies as a
function of their relative elliptical component $\hat{e}$ (see eqn.
[\ref{eq:ehat}]). For this figure, and the rest of this section, we
use the three template model for the reasons discussed in
\S~\ref{ssec:photoz_res}. There are two components -- one peak
consists of nearly pure Elliptical galaxies, while the other covers a
broad range of star forming galaxies. The spike at $\hat{e}=0$ is a
consequence of requiring $a_k\geq 0$ and contains $\sim$ 20\% of the
objects. If we allow negative spectral coefficients, the distribution
develops a smooth tail (see Fig. \ref{fg:type_hist}) where the best
fitting SEDs subtract an Elliptical component from the Sbc/Im
templates to reduce the 1.6$\mu$m emission peak. We inspected spectra
of galaxies in the spike, and found that it is dominated by galaxies
with obvious evidence of star formation, plus a number of galaxies
with photometry issues ($\sim 10\%$). If we modify the templates by
subtracting $20\%$ of the Elliptical template from the two star
forming ones, the spike shrinks, but at the cost of making the Sbc
template unphysical and worsening the photometric redshifts. If we
modify the components in this way and then refit the templates to the
AGES sample using the method of \S~\ref{ssec:temp_proc}, they move
back toward the original solution. A possible reason for this is the
lack of a second parameter such as metallicity to differentiate
between star forming galaxies, as the templates will converge to the
typical galaxies. Lack of data also contributes to the formation of
the spike. If we restrict the sample to galaxies with six or more
bands of photometry, the fraction of galaxies in the spike goes down
by about 20\%. Photometric redshift errors also contribute, since the
spike drops by 20\% if we use spectroscopic redshifts and by 50\% if
we apply the $\chi^2$ cuts of \S~\ref{ssec:photoz_res} (see
Fig. \ref{fg:type_hist}).

Figure \ref{fg:lum_shades} shows the distribution of the Elliptical
template fraction $\hat{e}$ of the SEDs as a function of their
bolometric luminosity $L$ for three different redshift ranges. Note
that while we use the 6\farcs0 aperture fluxes to fit the templates,
we correct to the total flux for the bolometric luminosity. We do this
by scaling the best-fit SED by the ratio of the Kron-like SExtractor $I$
band magnitude to the 6\farcs0 one. In practice however, the $I$-band
Kron-like photometry is sometimes affected by nearby bright stars
beyond the AGES flagging, producing excessively bright magnitudes. To
deal with this, we follow the approach of \citet{eisenstein07} and use
a total $I$-band magnitude produced from a weighted mean between the
Kron-like measurement and the predicted one from the $R$-band
measurement and the 6\farcs0 colors that favors the faintest of the
two. To account for the volume and depth limitations of the survey, we
use the $V/V_{\rm max}$ method \citep{schmidt68}, to properly weight
each bin for the effects of the magnitude limits. The density of each
bin is given by
\begin{equation}
n\ =\ \sum_i \frac{1}{\rm min[V_{\rm max}^i,V_{\rm max}] - V_{\rm
min}},
\end{equation}
\noindent where $V_{\rm max}^i$ is the co-moving volume to which
galaxy $i$ can be detected and $V_{\rm max}$ and $V_{\rm min}$ are the
volumes corresponding to the upper and lower edges of the redshift
bin. The $V_{\rm max}^i$ are easily calculated with our algorithm,
since it only depends on the SED of the galaxy and the magnitude
limits of the survey. In this figure we do not show the galaxies in
the $\hat{e} = 0$ spike. As expected, we see a well-defined clump of
galaxies with low star formation rate in all three redshift ranges,
and also a less well-defined locus of star forming galaxies that spans
a broader luminosity range. The latter group becomes less well-defined
at higher redshifts. Notice that in the lowest redshift bin
we cannot map the high star formation peak well mostly because of the
problems discussed above with the blue spike of Figure
\ref{fg:type_hist}.

Using the $V/V_{\rm max}$ method we also estimated B-band luminosity
functions for the NDWFS Bo\"otes field. We limited the survey to a
central area of approximately 5.3 deg$^2$ that is uniformly sampled in
all bands and contains $\sim 43000$ galaxies to $I < 21$. This is a
conservative limit of the usable survey area, but it will not affect
our results. As we did before, we excluded galaxies with a possible
AGN contamination (1200 objects), photometry in fewer than 4 bands
(1350) and near bright stars (2400), leaving us with a sample of
approximately 39000 galaxies. Using our templates and algorithms, we
predict the B-band \citep{bessell90} absolute magnitude of each
galaxy. We corrected for the dropped objects as a simple sampling
fraction correction. Naively implemented, the low luminosity tail of
the LF estimates are dominated by $L\sim L_*$ objects with
``catastrophic'' photo-z errors.  As discussed in section
\ref{ssec:photoz_res}, a $\chi^2$ cut can be implemented to minimize
such systematic failures. After some iterations, we decided on
eliminating the worst 10\% of the objects,leaving the sample with
$\sim 35000$ galaxies. This cut provides a good balance between
minimizing the statistical uncertainties from the diminished number of
objects and the systematic uncertainties from the photometric
redshifts. Our estimated luminosity functions, scaled to the total
number of galaxies in the selected sub-field (that is the ones used
for the estimation plus the ones with bad fits and the ones near
bright stars), are shown in Figure \ref{fg:lum_funcs} for four
redshift ranges and for four spectral type subdivisions: $0 \leq
\hat{e} < 0.4$ (high star formation rate ), $0.4 < \hat{e} < 0.8$
(intermediate star formation rate), $0.8 < \hat{e} \leq 1.0$ (low star
formation rate) and $0.0 \leq \hat{e} \leq 1.0$ (all star formation
rates). We estimated the errors by bootstrap re-sampling. This figure
also shows the best fit Schechter functions \citep{schech76} for each
case, with the parameters summarized in Table
\ref{tab:schech_fits}. We fit the Schechter functions only over the
magnitude ranges where the functional form is appropriate, dropping
the bins affected by the catalog magnitude limit and regions where
there is an apparent upturn at faint magnitudes. This upturn is
probably produced by a small artifact amplified by the $1/V_{max}$
weights. These present results are also limited by the photometry,
with problems in the total (Kron) magnitudes for objects with bright
neighbors affecting mostly the bright ends.

\citet{brown07} estimated the B-band luminosity functions of red
galaxies in the NDWFS field. The left panels of Figure
\ref{fg:early_lum_funcs} shows their results compared to our $0.8 <
\hat{e} \leq 1.0$ sample. Due to the different ways in which the
samples were selected, we only expect them to agree on the bright end
but not on the faint end slope or in the overall amplitude
$\phi_*$. \citet{brown07} defined their sample using the evolving and
luminosity dependent rest frame U--V color criterion (eqn. [3] of
\citealt{brown07})
\begin{equation}\label{eq:brown_crit}
\rm U - \rm V > 1.40 - 0.25 - 0.08(M_{\rm V} - 5 \log h + 20.0) -
0.42(z-0.05) + 0.07(z-0.05)^2 ,
\end{equation}
\noindent which corresponds to the expected location of the red
sequence in the $M_{\rm V}$ U--V plane displaced to the blue by 0.25
mag. Our criteria, on the other hand, corresponds to a non-evolving
and luminosity independent U--V color (U--V $\gtrsim 1.1$), so, by
definition, our sample will include fewer faint galaxies than
\citet{brown07}. Moreover, the evolution of the criteria set by
\citet{brown07} follows the evolution of the red sequence, becoming
bluer with increasing redshift, so we expect the differences between
the two luminosity functions to occur at brighter magnitudes at higher
redshifts, as seen in Figure \ref{fg:early_lum_funcs}.

\citet{wolf03} carried out a similar analysis to ours, using galaxies
from COMBO-17 with photometric redshifts and classifying them by their
overall spectral shape rather than their colors. In particular, their
{\it{type 1}} sample, defined as all galaxies with spectral types from
Ellipticals to Sab spirals, is similar to our low star formation rate
sample. We recalculated the luminosity functions using the COMBO-17
survey B-band for our early type sample, again keeping $\alpha$ fixed
to the value from the $0.2<z<0.4$ redshifts bin, and found in general
a good agreement with \citet{wolf03}. The right panels of Figure
\ref{fg:early_lum_funcs} show our luminosity functions compared to
those of \citet{wolf03} in the three redshift ranges where we
overlap. The agreement is very good for the two lowest redshift ranges
in the Figure, but somewhat worse for the highest one, although
still compatible.  A comparison with the rest of their results is not
straightforward, as there is no trivial match between their selection
criteria and ours for groups other than their {\it{type 1}}.

\section{Conclusions}

\begin{figure}
  \begin{center}
    \plotone{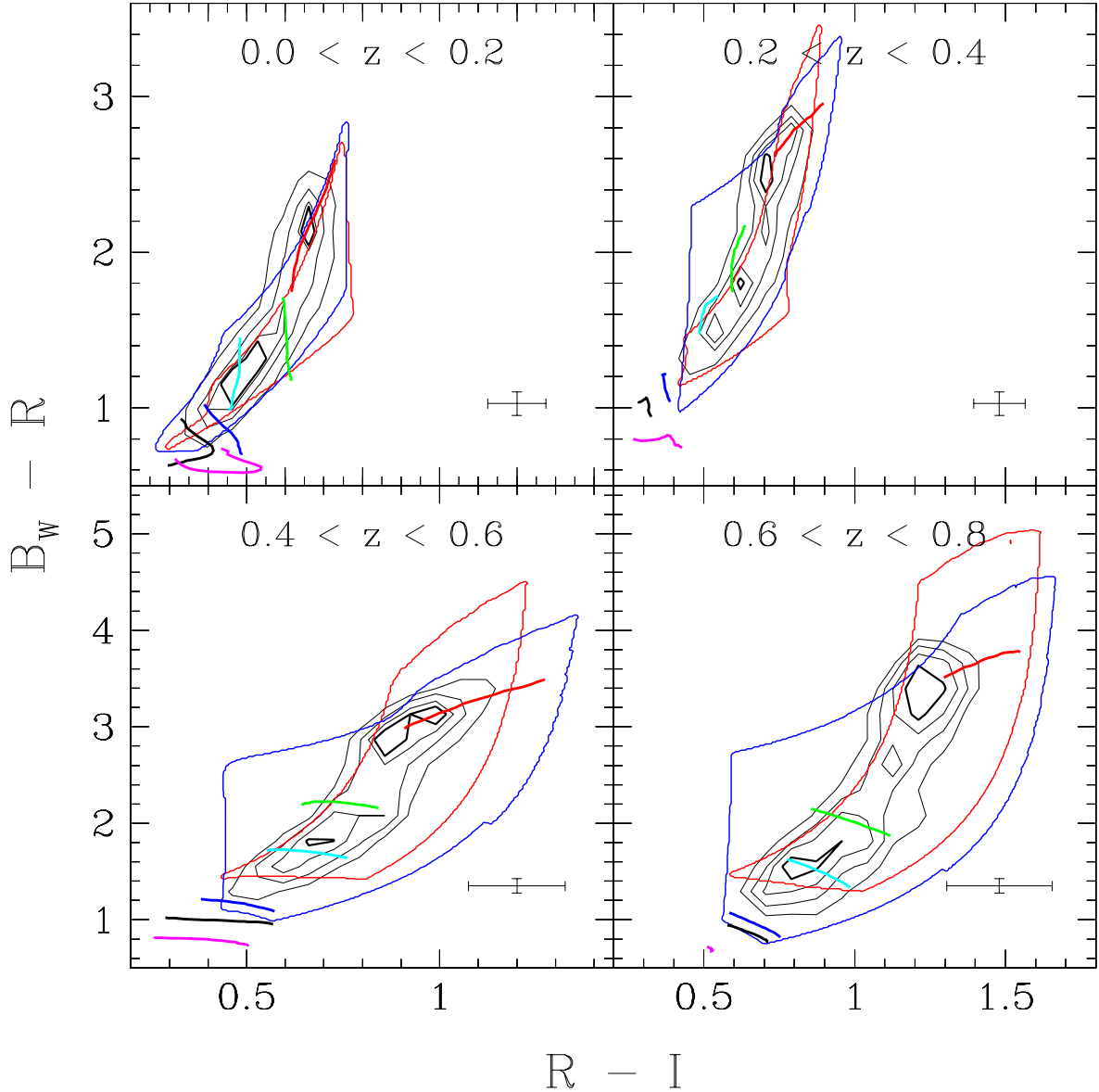}
    \caption{The color--color distributions of the AGES galaxy sample
    for four different redshift ranges in the optical bands. The black
    contours enclose 20 (bold), 40, 60 and 80\% of the galaxies in the
    sample. Solid lines mark the borders of the areas covered by our
    three {\it{(red)}} and four {\it{(blue)}} template models. The
    error bars in the bottom right of each panel show the typical
    color error for galaxies in each sample. For comparison we show in
    thick lines the colors of six common templates from the
    literature: CWW Elliptical {\it{(red)}}, Sbc {\it{(green)}}, Scd
    {\it{(cyan)}} and Im {\it{(blue)}}, and \citet{kinney96} SB1
    {\it{(yellow)}} and SB2 {\it{(magenta)}}.}
    \label{fg:color_diag_opt}
  \end{center}
\end{figure}

\begin{figure}
  \begin{center}
    \plotone{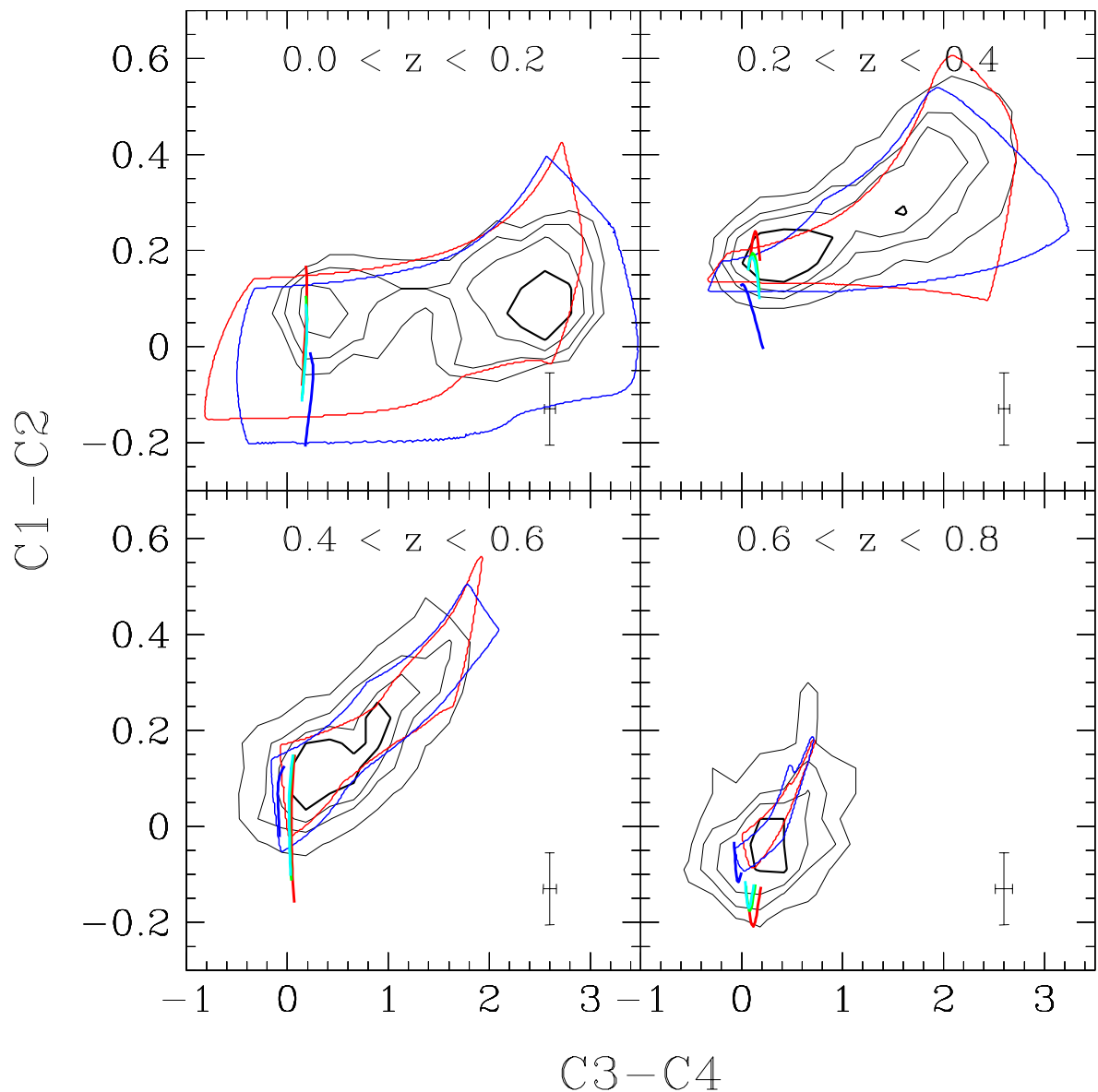}
    \caption{The color--color distributions of the AGES galaxy sample
    for four different redshift ranges in the mid-IR bands. Contours
    are defined in the same way as in
    Fig. \ref{fg:color_diag_opt}. For comparison, we show the
    \citet{bc03} extended CWW templates in the same color-coding as in
    the optical. Notice that the E, Sbc and Scd colors sometimes
    overlap since they are very similar in this wavelength range. Also
    note that the low redshift galaxy distribution is strongly
    bimodal.}
    \label{fg:color_diag_mir}
  \end{center}
\end{figure}

\begin{figure}
  \begin{center}
    \includegraphics[width=0.6\textwidth]{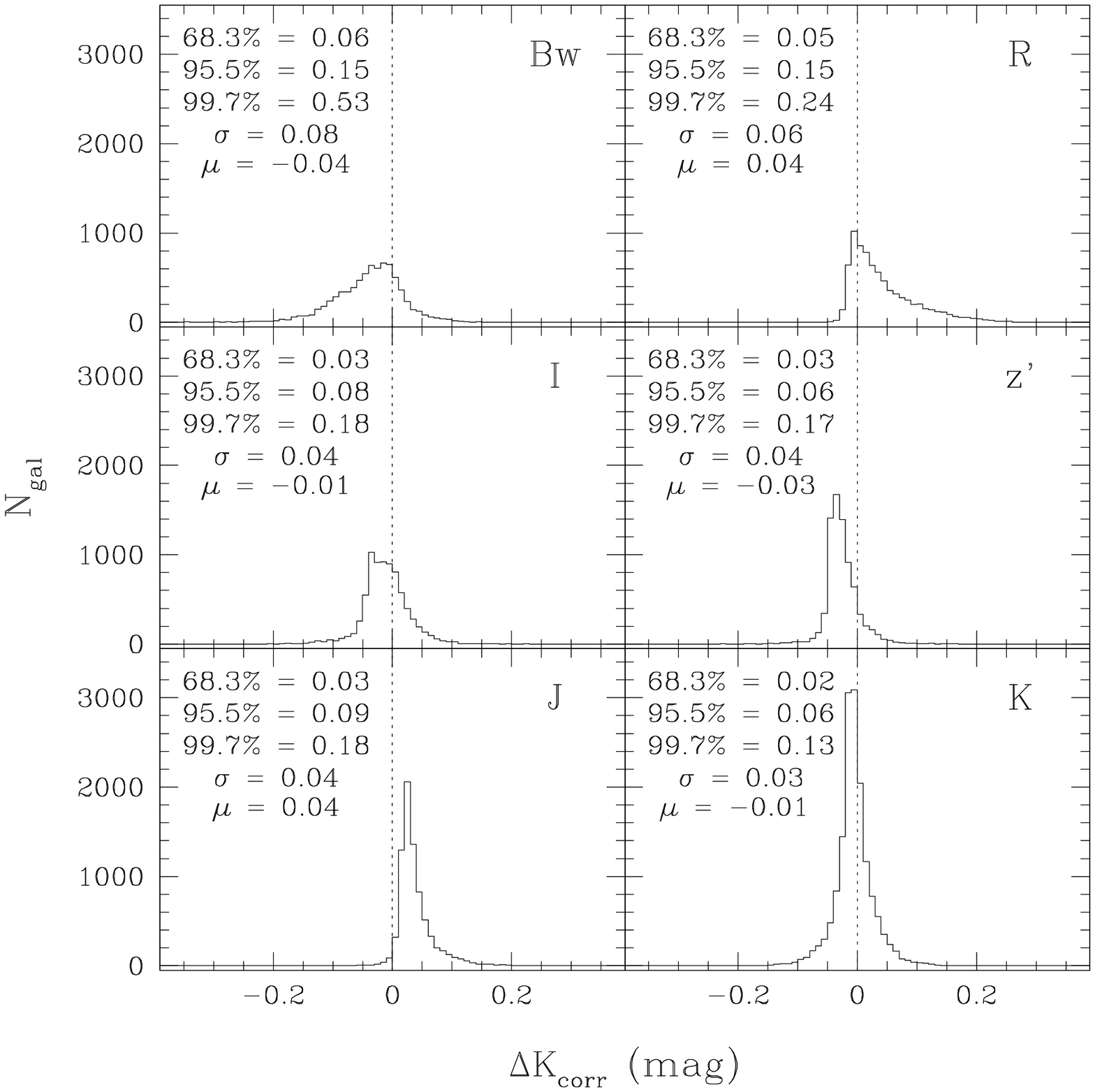}
    \includegraphics[width=0.6\textwidth]{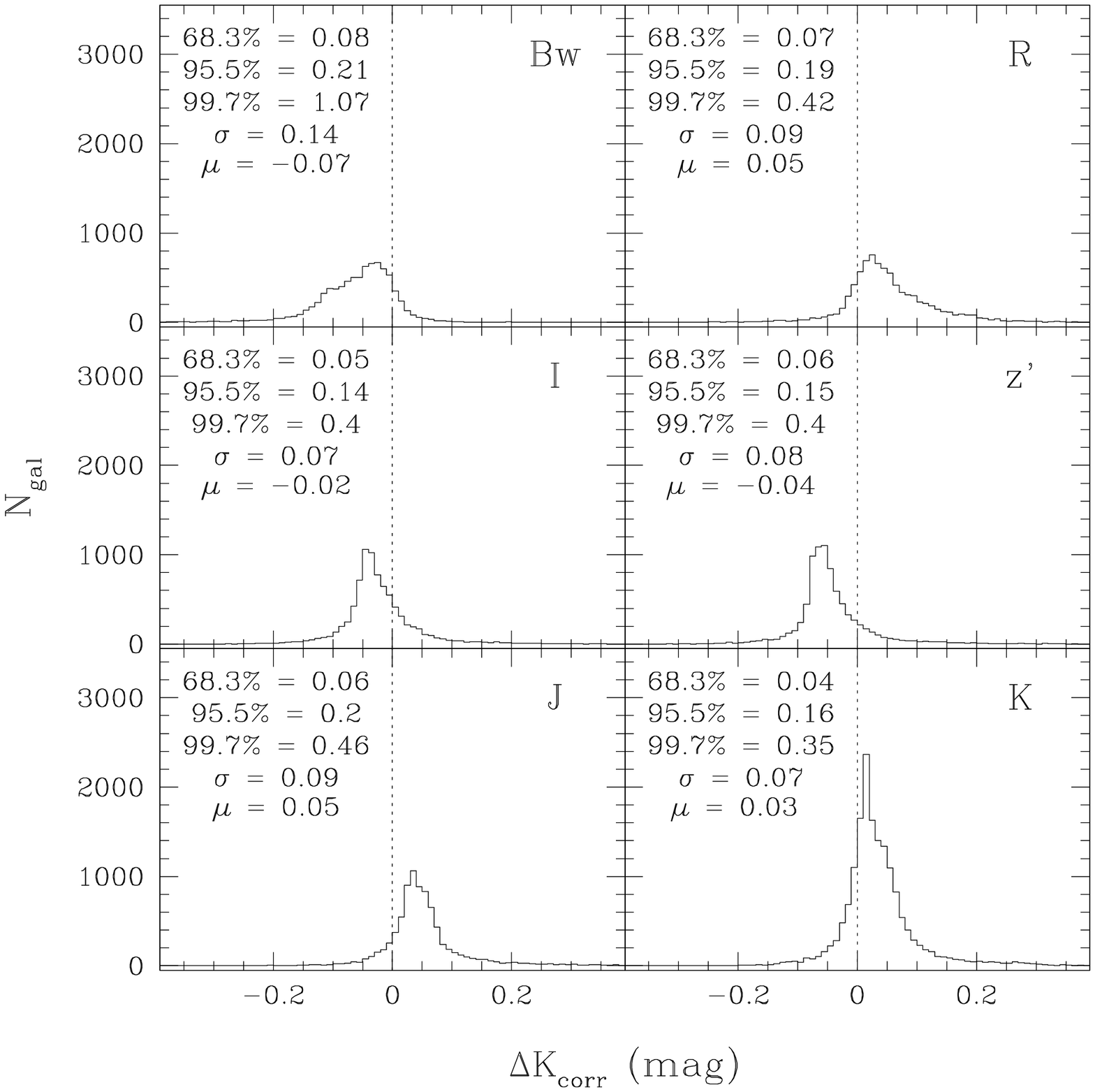}
    \caption{Histograms of the differences between the $K$ corrections
    determined here and those determined by {\tt{kcorrect v4\_1\_4}}
    of \citet{blanton03} for the AGES galaxy sample in all optical and
    near-IR AGES bands (K and K$_{\rm s}$ have been combined into a
    single K band) for redshifts lower {\textit{(top)}} and higher
    {\textit{(bottom)}} than 0.3. Each panel also gives the standard
    deviation between the methods $\sigma$, the mean $\mu$ of $\Delta
    K_{corr}$ and the values of $\left|\Delta K_{corr}\right|$ that
    encompasses 68.3, 95.5 and 99.7\% of the objects. The IRAC bands
    are not considered since {\tt{kcorrect v4\_1\_4}} cannot model
    mid-IR fluxes.}
    \label{fg:kcorr}
  \end{center}
\end{figure}

\begin{figure}
  \begin{center}
    \plotone{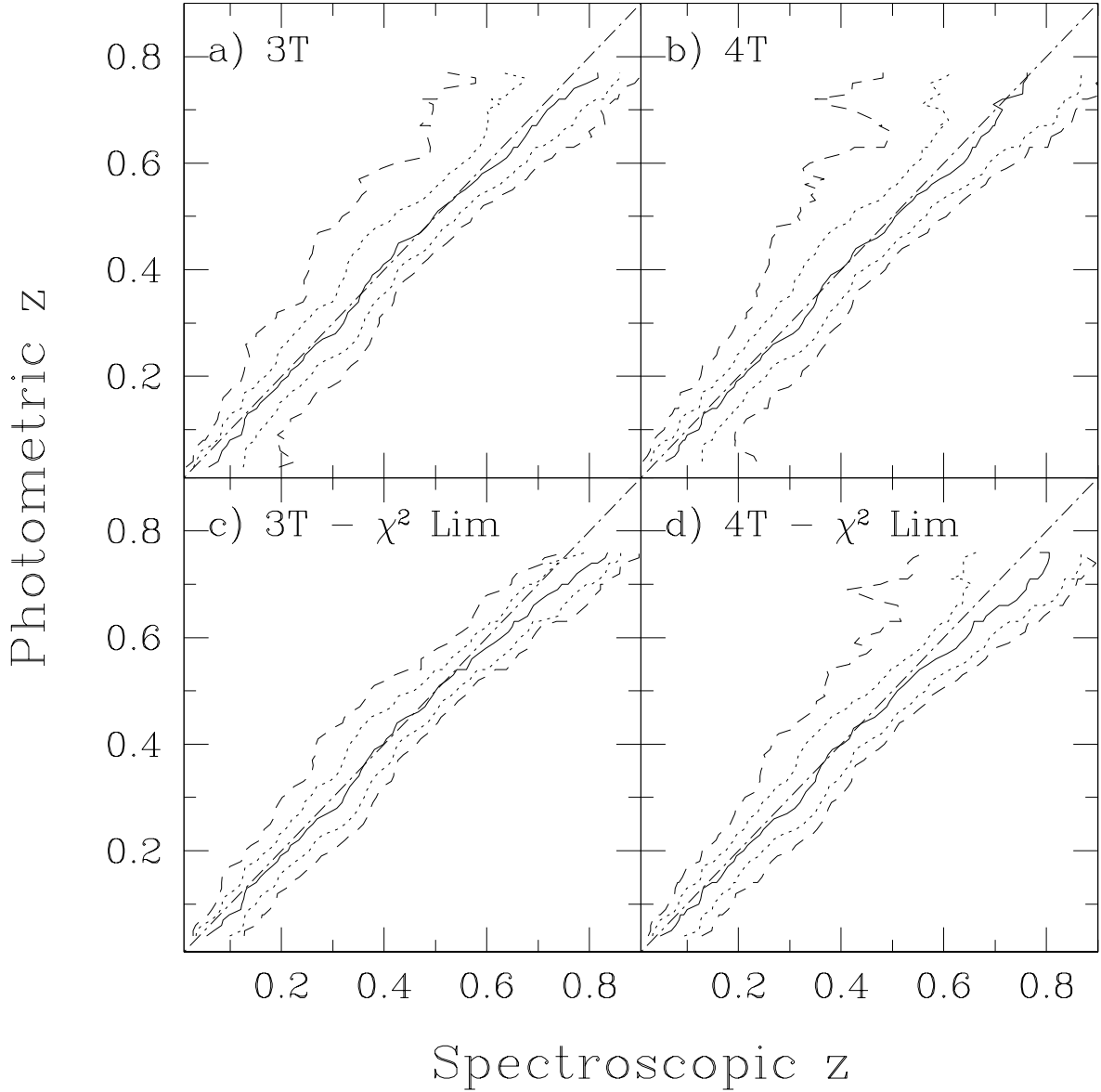}
    \caption{Comparison of photometric and spectroscopic
    redshifts. For a fixed photometric redshift, the solid line shows
    the median of the spectroscopic redshifts, while the dotted and
    dashed lines contain the 68.3 and 90\% of the distribution
    respectively. The two redshifts are equal, $z_p = z_s$, on the
    diagonal dot dashed line. Panel $a)$ shows the comparison for the
    3 template model and $b)$ for the 4 template one. For the bottom
    panels, $c)$ and $d)$, we have only included the 75\% of the
    objects for which there is a probability greater than 10\% of
    obtaining a $\chi^2$ larger than that of the best fit.}
    \label{fg:zphot}
  \end{center}
\end{figure}

\begin{figure}
  \begin{center}
    \plotone{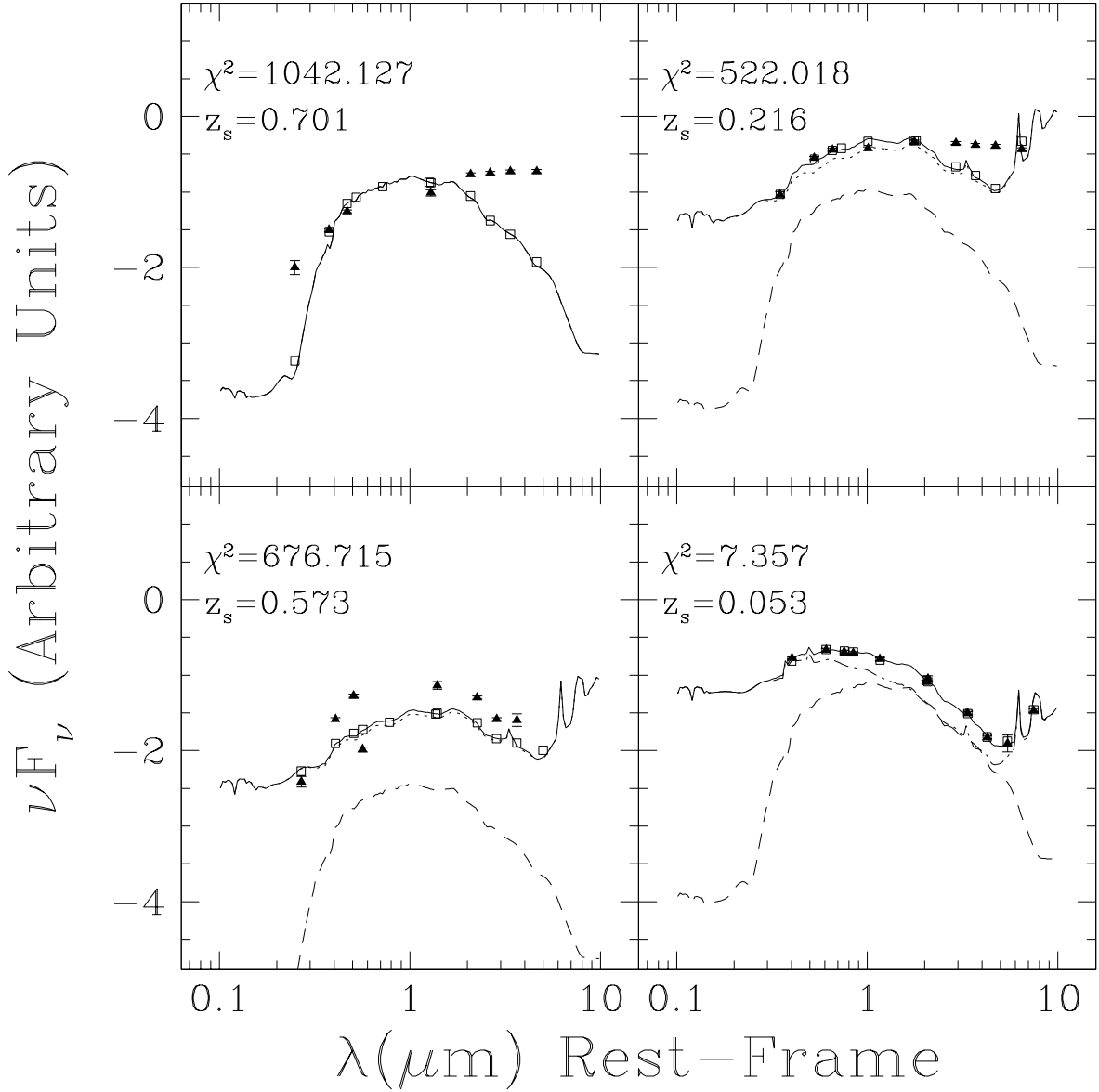}
    \caption{Examples of bad and good fits to the data. In each panel
    we show the photometric data {\it{(triangles)}}, the model
    bandpass fluxes {\it{(squares)}}, the overall model SED
    {\it{(solid line)}} and the E {\it{(dashed line)}}, Sbc
    {\it{(dotted line)}} and Im {\it{(dot-dashed line)}} contributions
    to the model SED. The top panels show the fit for galaxies with
    AGN contamination. The bottom left panel is a galaxy with bad
    photometry in the z' band. Finally, the bottom right panel shows
    the median fit for comparison.}
    \label{fg:bad_ex}
  \end{center}
\end{figure}

\begin{figure}
  \begin{center}
    \plotone{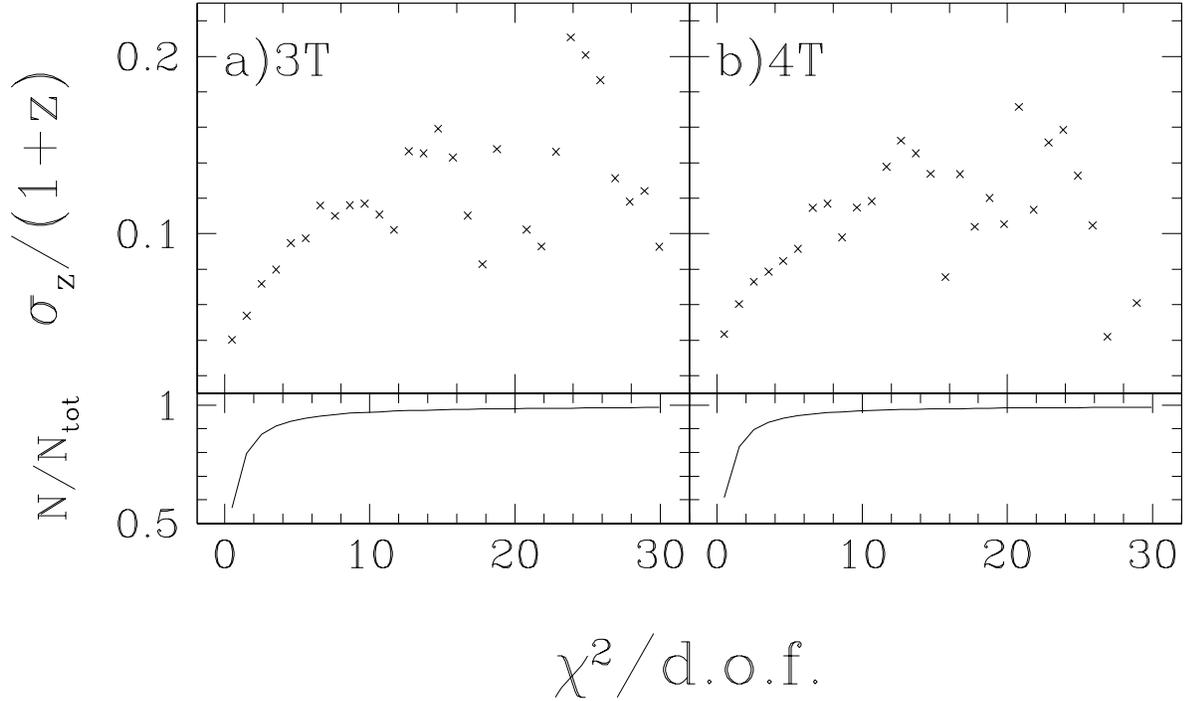}
    \caption{{\it{(Top)}} The dispersion $\sigma_z/(1+z)$ defined in
    equation (\ref{eq:sigmaz}) as a function of the $\chi^2$ per
    degree of freedom for the fits to the photometry at the best
    photometric redshift in the three {\it{(left)}} and four
    {\it{(right)}} template models. The points are the mean values for
    objects divided in bins with a width of one unit of $\chi^2$ per
    degree of freedom. {\it{(Bottom)}} The fraction of objects with
    fits better than that $\chi^2/N_{\rm DOF}$. The correlation
    between $\chi^2$ and $\sigma_z/(1+z)$ justifies the $\chi^2$ cut
    used in the bottom panel of Figure \ref{fg:zphot}.}
    \label{fg:chi2z}
  \end{center}
\end{figure}

\begin{figure}
  \begin{center}
    \plotone{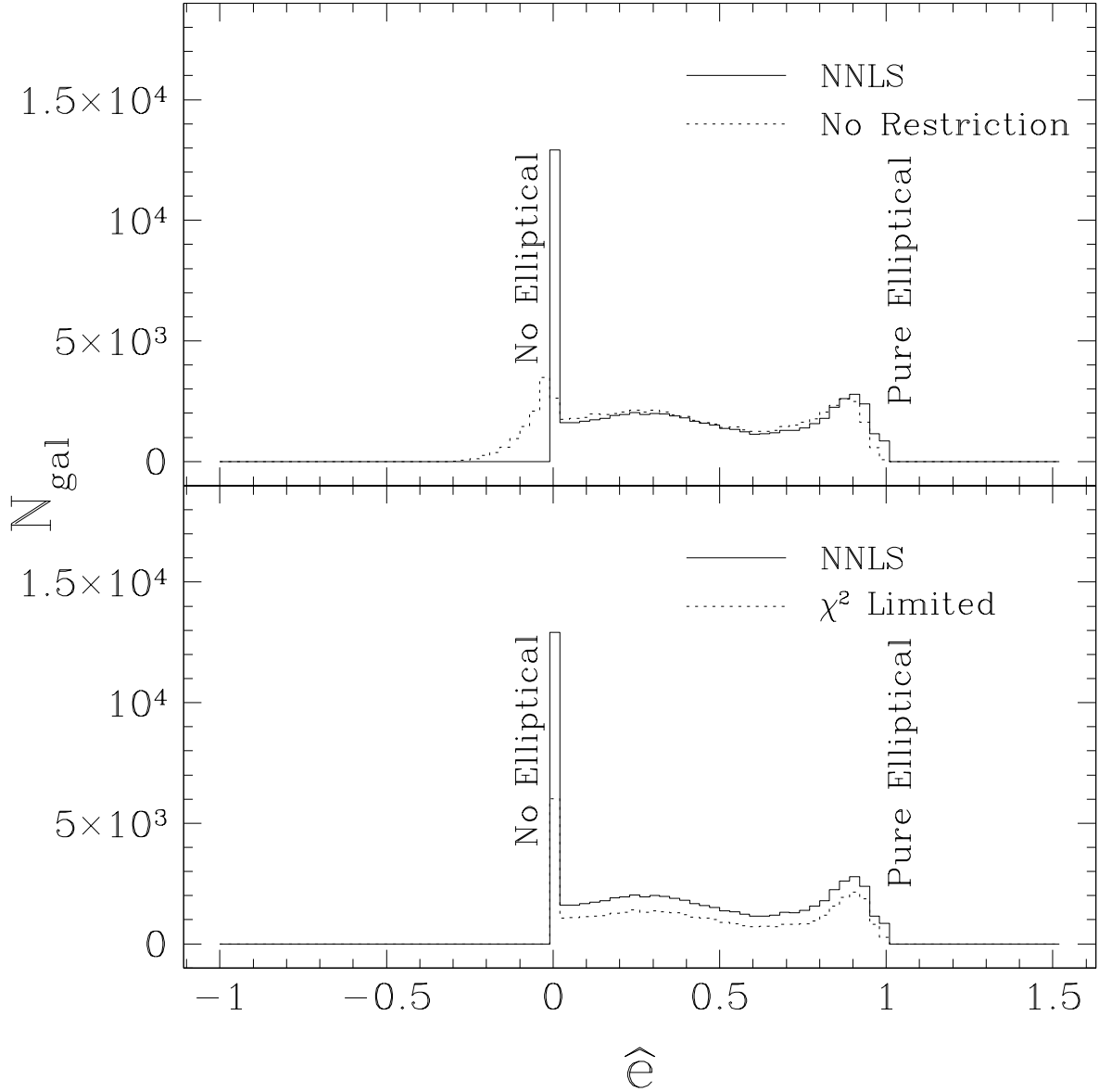}
    \caption{Distribution of galaxies as a function of the elliptical
    component fraction $\hat{e}$ in their SED. In both panels, the
    bold solid line shows the distribution obtained using the NNLS
    algorithm to enforce $a_k\geq 0$. The dotted line in the top panel
    shows the distribution when we drop this restriction, while in the
    bottom panel it shows the distribution when using the NNLS
    algorithm but applying the $\chi^2$ cut of
    \S~\ref{ssec:photoz_res}.}
    \label{fg:type_hist}
  \end{center}
\end{figure}

\begin{figure}
  \begin{center}
    \plotone{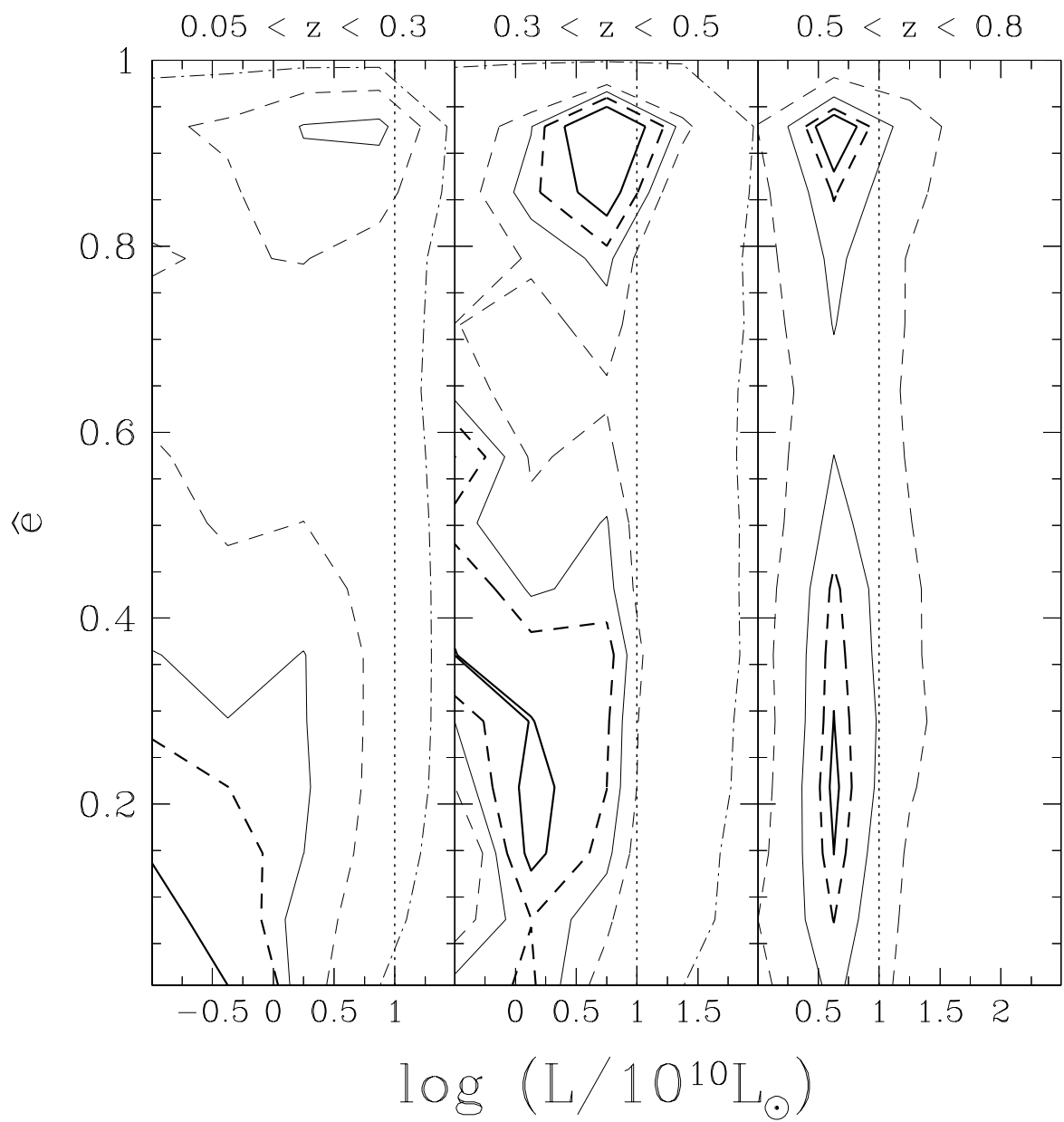}
    \caption{Contour plots of the distribution of galaxies as a
    function of their bolometric luminosity and the amount of
    elliptical component in their SEDs for three redshift ranges.
    The contour levels, with the ordering of bold solid, bold
    dashed, thin solid, thin dashed and thin dot-dashed, enclose
    20\%,40\%,60\% and 100\% of the objects respectively. The
    vertical dotted line shows $\log{L/10^{10} L_{\odot}} = 1$. In
    each redshift range we see a bimodal distribution of galaxies with
    either high or low star formation rates.}
    \label{fg:lum_shades}
  \end{center}
\end{figure}

We have built an optimized basis of low resolution spectral templates
for the wavelength range from 0.2--10 $\mu$m that accurately reproduce
most galaxy SEDs. We used a variant of the \citet{budava00} method to
fit the SEDs of 17000 AGES galaxies with photometry in at least 6 of
11 possible bands. We considered a three template basis starting from
the CWW Elliptical, Sbc and Im templates and a four template basis
where we added an E+A post-starburst component. One novel feature of
our approach is that we model each galaxy as a non-negative sum of
templates, which markedly improves the match of the model to the
observed color range of galaxies (see Figs. \ref{fg:color_diag_opt}
and \ref{fg:color_diag_mir} ) and significantly improves photometric
redshift estimates.

We applied these optimized templates to calculate accurate photometric
redshifts. We find that while the four template models fit the galaxy
SEDs better than the three template models when the redshift is known,
they broaden the photometric redshifts errors by approximately
50\%. Using the three templates basis, we showed that the accuracy of
our method is $\sigma_z/(1+z) = 0.060$ ($\Delta z = 0.038$), with the
accuracy being highest for early type galaxies. Many of the galaxies
with poor photometric redshifts estimates are also poorly fit by the
templates because of either bad photometric data points or AGN
contamination. If we consider only galaxies having $\chi^2$ values
smaller than the 90th percentile of their expected value, the accuracy
improves to $\sigma_z/(1+z) = 0.044$ ($\Delta z = 0.030$) when
dropping the worst 5\% of the redshifts. This is somewhat better than
that obtained by \citet{brodwin06} for a very similar data set but
using a hybrid approach that mixed SED fitting and neural
networks. Our results are somewhat worse than those obtained by the
ZEBRA code \citep{zebra06} for a COSMOS \citep{cosmos06} galaxy
sample, but this is probably due to the very small number of degrees
of freedom in their data set after fitting six redshift-dependent
templates to a sample of only 866 galaxies that is then used to test
those templates.

Besides photometric redshifts, we also applied these optimized
templates to calculate accurate $K$ corrections and bolometric
luminosities. We compared the $K$ corrections to those obtained using
the \verb+kcorrect v4_1_4+ code of \citet{blanton03} for the AGES
galaxy sample and found a very good agreement between them. We have
implemented our algorithms for calculating bolometric luminosities,
$K$ corrections and photometric redshifts, including our optimized
template basis, in a public code\footnote{Code available at
www.astronomy.ohio-state.edu/$\sim$rjassef/lrt} that can carry out the
calculations for any set of filters provided by the user.

\begin{figure}
  \begin{center}
    \plotone{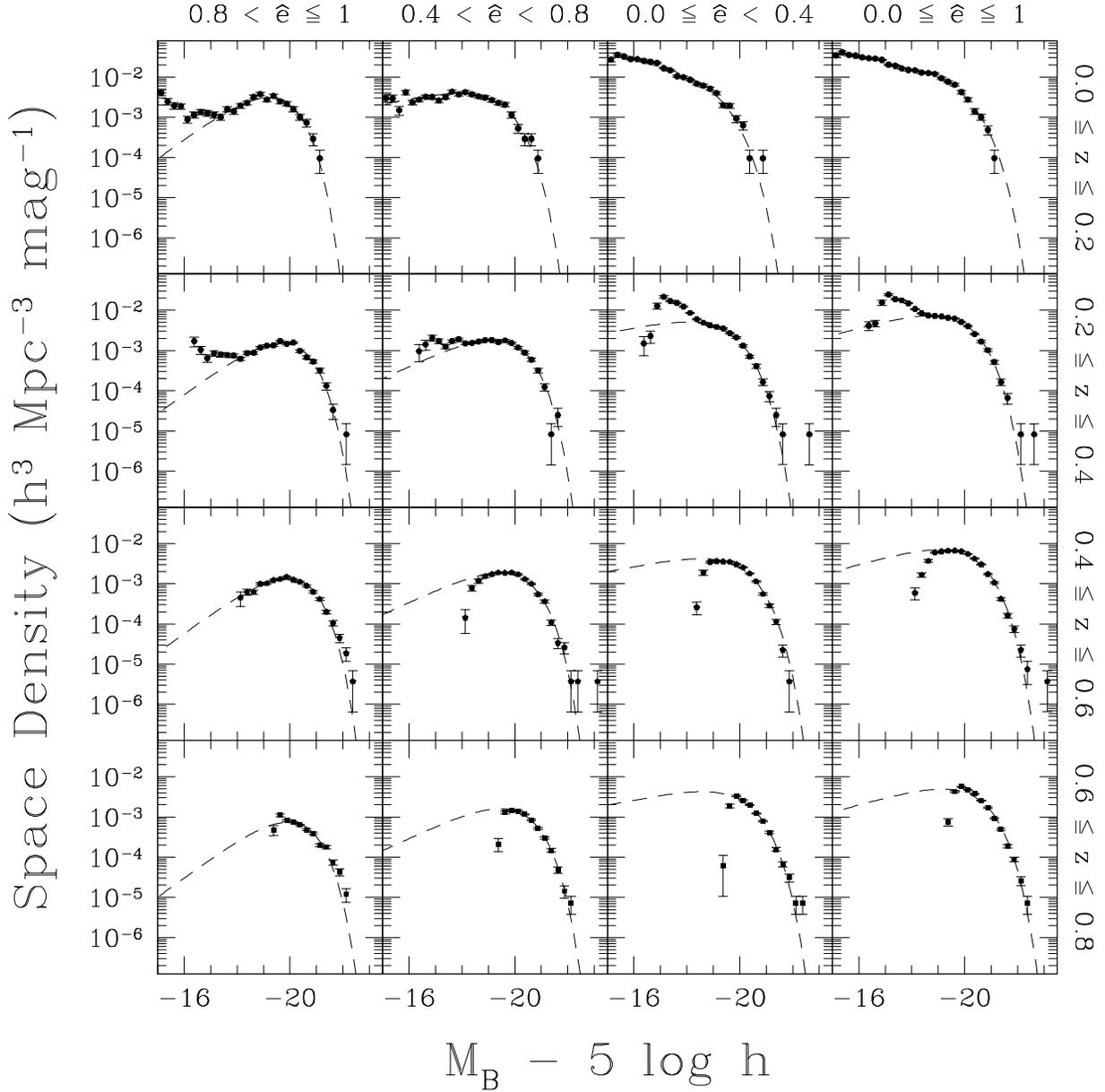}
    \caption{Luminosity functions for the NDWFS Bo\"otes field. They
    are divided into three $\hat{e}$ ranges 1--0.8 {\it{(left)}},
    0.8--0.4 {\it{(middle-left)}} and 0.4--0.0 {\it{(middle-right)}},
    and into four redshift ranges 0.0--0.2 {\it{(top)}}, 0.2--0.4
    {\it{(middle-top)}}, 0.4--0.6 {\it{(middle-bottom)}} and 0.6--0.8
    {\it{(bottom)}}. We also show the luminosity function of all
    galaxies for the same redshift ranges in the rightmost panel. The
    dashed lines show the best fit Schechter function. For all
    $\hat{e}$ ranges there seems to be an evolutionary trend with
    redshift.}
    \label{fg:lum_funcs}
  \end{center}
\end{figure}

\begin{figure}
  \begin{center}
    \plotone{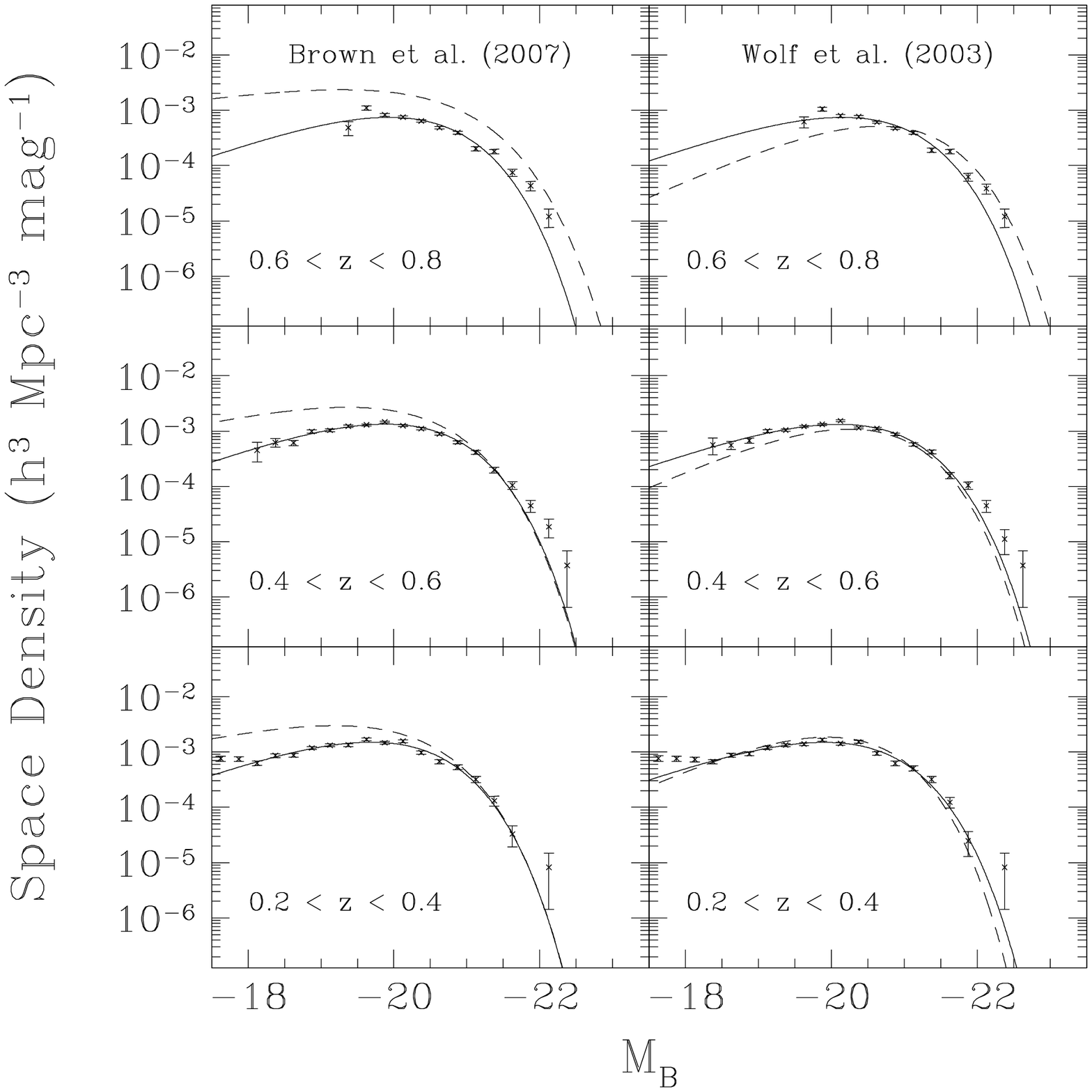}
    \caption{Early-type galaxy luminosity functions for the redshift
    ranges 0.2--0.4 {\it{(bottom)}}, 0.4--0.6 {\it{(middle)}} and
    0.6--0.8 {\it{(top)}} from this work {\it{(solid lines and
    points)}} and from \citet[][left]{brown07} and
    \citet[][right]{wolf03} {\it{(dashed lines)}}. Note that the
    shapes agree well for galaxies brighter than $M_*$. The difference
    in the fainter end and in the normalization $\phi_*$ with the
    functions of \citet{brown07} come from the different selection
    criteria (see \S~\ref{sec:spec_clas} for details).}
    \label{fg:early_lum_funcs}
  \end{center}
\end{figure}

We applied these algorithms to the photometric galaxy sample of the
NDWFS Bo\"otes field with $I \leq 21$ mag ($\sim 69000$ galaxies)
and studied the galaxy luminosity distribution as a function of
redshift and star formation (parametrized by the early-type template
fraction $\hat{e}$). We find that our algorithms reproduce the bimodal
distribution of red and blue galaxies that has been observed as a
function of color and magnitude in the SDSS
\citep{strateva01,blanton03b,kauffmann03}, DEEP2
\citep{madgwick03,weiner05} and COMBO-17 \citep{bell04} surveys, for
example. We have also shown that the mid-infrared color-color
distribution of galaxies is strongly bimodal, resembling its optical
counterpart, except that rather than a red clump and a blue cloud, it
has a blue clump and a red cloud
(Fig. \ref{fg:color_diag_mir}). Finally, we used these algorithms to
estimate the B-band luminosity functions of the field from a central
region of the survey containing about 43000 galaxies. Our approach
allows us to easily study them as a function of redshift and star
formation. Our results, summarized in Figure \ref{fg:lum_funcs} and
Table \ref{tab:schech_fits}, agree broadly with the results of
\citet{brown07} and \citet{wolf03}.

\acknowledgments

We wish thank Richard W. Pogge for lending us his expertize in
analyzing spectra, Steve Willner for his suggestions and comments, and
all the people in the NDWFS, FLAMEX and IRAC Shallow Survey
collaborations that did not directly participate in this work. The
AGES observations were obtained at the MMT Observatory, a joint
facility of the Smithsonian Institution and the University of
Arizona. This work made use of data products provided by the NOAO Deep
Wide-Field Survey \citep{ndwfs99,jannuzi05,dey05}, which is supported
by the National Optical Astronomy Observatory (NOAO). This research
draws upon data provided by Dr. Buell Jannuzi and Dr. Arjun Dey as
distributed by the NOAO Science Archive. NOAO is operated by AURA,
Inc., under a cooperative agreement with the National Science
Foundation. This work is based in part on observations made with the
Spitzer Space Telescope, which is operated by the Jet Propulsion
Laboratory, California Institute of Technology under a contract with
NASA.

\include{tab1}

\include{tab2}

\include{tab3}
\include{tab4}
\include{tab5}
\include{tab6}

\end{document}

%% file: tab1.tex
\begin{deluxetable}{c c c c}

\tablehead{$\lambda$($\mu$m) & \multicolumn{3}{c}{$F_{\nu}$ ($\times10^{-14}$ erg/s/cm$^2$/Hz)}\\  & E & Sbc & Im}

\tablecaption{Three Component Spectral Templates \label{tab:3spectab}}
\tabletypesize{\small}
\tablewidth{0pt}
\tablecolumns{4}

\startdata

    0.1000 &     0.1086 &     3.8414 &    18.6522 \\
    0.1029 &     0.1239 &     4.8336 &    23.4179 \\
    0.1059 &     0.1220 &     4.8676 &    23.4784 \\
    0.1090 &     0.1267 &     5.2671 &    25.1332 \\
    0.1122 &     0.1267 &     5.3591 &    25.2659 \\

\enddata

\tablecomments{Electronic table that presents the flux per unit
frequency $F_{\nu}$ of the three components model best fit template
spectra as a function of wavelength. Templates are normalized to be at
a distance of 10pc and to have an integrated luminosity between the
wavelength boundaries of $10^{10} L_{\odot}$.}
\end{deluxetable}

%% file: tab2.tex
\begin{deluxetable}{c c c c c}

\tablehead{$\lambda$($\mu$m) & \multicolumn{4}{c}{$F_{\nu}$ ($\times10^{-14}$ erg/s/cm$^2$/Hz)}\\ & E & Sbc & Im & E+A}

\tablecaption{Four Component Spectral Templates \label{tab:4spectab}}
\tabletypesize{\small}
\tablewidth{0pt}
\tablecolumns{5}

\startdata

    0.1000 &     0.1430 &     3.5994 &    27.7314 &     0.0822 \\
    0.1029 &     0.1632 &     4.5298 &    34.8230 &     0.0942 \\
    0.1059 &     0.1604 &     4.5628 &    34.9195 &     0.0929 \\
    0.1090 &     0.1670 &     4.9381 &    37.3873 &     0.1383 \\
    0.1122 &     0.1665 &     5.0253 &    37.5922 &     0.1935 \\

\enddata

\tablecomments{Electronic table that presents the flux per unit frequency
$F_{\nu}$ of the four components model best fit template spectra as a
function of wavelength. Templates are normalized to be at a distance
of 10pc and to have an integrated luminosity between the wavelength
boundaries of $10^{10} L_{\odot}$.}
\end{deluxetable}

%% file: tab3.tex
\begin{deluxetable}{c c c c c c c c c c c c c c c c c c c c c}

\tablehead{$z$ & Template & B$_{\rm w}$ & B & V & R & I & u' & g' & r' & i' & z' & J & H & K$_{\rm s}$ & K & C1 & C2 & C3 & C4 &DM}

\rotate
\setlength{\tabcolsep}{0.03in} 
\tablecaption{Three Template Model Absolute Magnitudes\label{tab:3s_modmag}}
\tabletypesize{\scriptsize}
\tablewidth{0pt}
\tablecolumns{21}

\startdata

0.0	 &1	 &	 --18.47  &	 --18.57  &	 --19.49  &	 --20.14  &	 --20.82  &	 --17.03  &	 --18.97  &	 --19.78  &	 --20.20  &	 --20.57  &	 --21.68  &	 --22.49  &	 --22.66  &	 --22.63  &	 --22.99  &	 --22.84  &	 --22.81  &	 --22.00  &           \\
0.0	 &2	 &	 --17.96  &	 --17.99  &	 --18.61  &	 --19.29  &	 --19.90  &	 --17.20  &	 --18.25  &	 --18.94  &	 --19.30  &	 --19.70  &	 --21.00  &	 --21.94  &	 --22.36  &	 --22.35  &	 --23.29  &	 --23.44  &	 --25.39  &	 --27.40  &           \\
0.0	 &3	 &	 --19.34  &	 --19.35  &	 --19.72  &	 --20.11  &	 --20.40  &	 --18.78  &	 --19.56  &	 --19.85  &	 --19.92  &	 --20.02  &	 --20.82  &	 --21.36  &	 --21.49  &	 --21.47  &	 --21.93  &	 --21.94  &	 --23.67  &	 --25.31  &           \\
0.1	 &1	 &	 --17.89  &	 --18.06  &	 --19.32  &	 --20.02  &	 --20.73  &	 --16.41  &	 --18.64  &	 --19.66  &	 --20.10  &	 --20.50  &	 --21.77  &	 --22.47  &	 --22.88  &	 --22.88  &	 --23.18  &	 --23.20  &	 --23.18  &	 --22.53  &      38.32\\
0.1	 &2	 &	 --17.70  &	 --17.74  &	 --18.51  &	 --19.16  &	 --19.85  &	 --17.09  &	 --18.11  &	 --18.76  &	 --19.28  &	 --19.58  &	 --21.04  &	 --21.85  &	 --22.55  &	 --22.56  &	 --23.42  &	 --23.59  &	 --24.39  &	 --27.28  &      38.32\\
0.1	 &3	 &	 --19.21  &	 --19.23  &	 --19.77  &	 --20.11  &	 --20.50  &	 --18.57  &	 --19.49  &	 --19.83  &	 --20.02  &	 --20.07  &	 --20.95  &	 --21.48  &	 --21.71  &	 --21.71  &	 --22.19  &	 --22.16  &	 --22.75  &	 --25.34  &      38.32\\
0.2	 &1	 &	 --17.20  &	 --17.50  &	 --19.12  &	 --19.89  &	 --20.64  &	 --15.53  &	 --18.16  &	 --19.50  &	 --20.01  &	 --20.42  &	 --21.82  &	 --22.45  &	 --23.11  &	 --23.11  &	 --23.36  &	 --23.50  &	 --23.43  &	 --23.10  &      39.96\\
0.2	 &2	 &	 --17.42  &	 --17.47  &	 --18.44  &	 --19.03  &	 --19.80  &	 --17.07  &	 --17.88  &	 --18.65  &	 --19.17  &	 --19.53  &	 --21.04  &	 --21.81  &	 --22.68  &	 --22.70  &	 --23.37  &	 --23.79  &	 --24.20  &	 --26.92  &      39.96\\
0.2	 &3	 &	 --18.99  &	 --19.01  &	 --19.74  &	 --20.15  &	 --20.58  &	 --18.51  &	 --19.37  &	 --19.87  &	 --20.07  &	 --20.16  &	 --21.07  &	 --21.56  &	 --21.93  &	 --21.93  &	 --22.30  &	 --22.40  &	 --22.64  &	 --25.07  &      39.96\\
0.3	 &1	 &	 --16.71  &	 --16.95  &	 --18.77  &	 --19.75  &	 --20.55  &	 --14.47  &	 --17.60  &	 --19.33  &	 --19.92  &	 --20.34  &	 --21.79  &	 --22.51  &	 --23.25  &	 --23.26  &	 --23.50  &	 --23.69  &	 --23.68  &	 --23.56  &      40.96\\
0.3	 &2	 &	 --17.28  &	 --17.29  &	 --18.30  &	 --18.95  &	 --19.70  &	 --17.02  &	 --17.64  &	 --18.59  &	 --19.02  &	 --19.51  &	 --20.97  &	 --21.84  &	 --22.71  &	 --22.75  &	 --23.39  &	 --23.98  &	 --24.23  &	 --26.41  &      40.96\\
0.3	 &3	 &	 --18.76  &	 --18.82  &	 --19.70  &	 --20.15  &	 --20.60  &	 --18.43  &	 --19.19  &	 --19.89  &	 --20.04  &	 --20.24  &	 --21.15  &	 --21.64  &	 --22.10  &	 --22.11  &	 --22.44  &	 --22.65  &	 --22.74  &	 --24.65  &      40.96\\
\enddata

\tablecomments{\small{The electronic table supplies the absolute
magnitude of the three template model as a function of redshift,
along with the distance modulus DM. A complete version of this table
can be found in the electronic edition of the journal. The absolute
magnitude we present here corresponds to the canonical definition of
the absolute magnitude \citep[as in, for example, eqn. 26
of][]{hogg99} plus the $K$ correction term. This allows the
calculation of photometric redshifts and $K$ corrections from the
table. To determine photometric redshifts, colors should be calculated
and matched to the data by varying the $a_k$ coefficients (see
\S~\ref{ssec:temp_proc}) and the redshift. For a galaxy at redshift
$z$ with template coefficients $a_k$, the model magnitude in band $b$
is given by $M_b(z)\ =\ -2.5\ \log{\sum_k a_k 10^{-0.4
M_{b,k}(z)}}$. Apparent magnitudes can be determined by adding the
distance modulus to the absolute ones. To determine $K$ corrections
for a galaxy at redshift $z$, coefficients $a_k$ should also be
determined to match the observed colors as above. With the same
coefficients, redshift $z$ and redshift zero model absolute magnitudes
can be determined, and the difference between them will correspond to
the desired $K$ correction.}}
\end{deluxetable}

%% file: tab4.tex
\begin{deluxetable}{c c c c c c c c c c c c c c c c c c c c c}

\tablehead{$z$ & Template & B$_{\rm w}$ & B & V & R & I & u' & g' & r' & i' & z' & J & H & K$_{\rm s}$ & K & C1 & C2 & C3 & C4 &DM}

\rotate
\setlength{\tabcolsep}{0.03in} 
\tablecaption{Four Template Model Absolute Magnitudes\label{tab:4s_modmag}}
\tabletypesize{\scriptsize}
\tablewidth{0pt}
\tablecolumns{21}

\startdata

0.0	 &1	 &	 --18.29  &	 --18.41  &	 --19.45  &	 --20.11  &	 --20.81  &	 --16.86  &	 --18.87  &	 --19.74  &	 --20.19  &	 --20.57  &	 --21.73  &	 --22.52  &	 --22.73  &	 --22.70  &	 --23.00  &	 --22.80  &	 --22.69  &	 --22.32  &           \\
0.0	 &2	 &	 --18.09  &	 --18.13  &	 --18.78  &	 --19.46  &	 --20.06  &	 --17.15  &	 --18.40  &	 --19.12  &	 --19.46  &	 --19.86  &	 --21.12  &	 --22.09  &	 --22.49  &	 --22.48  &	 --23.36  &	 --23.53  &	 --25.37  &	 --27.17  &           \\
0.0	 &3	 &	 --18.87  &	 --18.86  &	 --19.23  &	 --19.62  &	 --19.99  &	 --18.49  &	 --19.08  &	 --19.35  &	 --19.47  &	 --19.67  &	 --20.61  &	 --21.28  &	 --21.53  &	 --21.50  &	 --22.16  &	 --22.07  &	 --24.28  &	 --26.63  &           \\
0.0	 &4	 &	 --19.68  &	 --19.70  &	 --20.04  &	 --20.43  &	 --20.70  &	 --18.69  &	 --19.87  &	 --20.18  &	 --20.22  &	 --20.29  &	 --21.07  &	 --21.58  &	 --21.79  &	 --21.76  &	 --22.22  &	 --22.35  &	 --23.08  &	 --23.77  &           \\
0.1	 &1	 &	 --17.65  &	 --17.85  &	 --19.27  &	 --19.99  &	 --20.71  &	 --16.33  &	 --18.47  &	 --19.62  &	 --20.07  &	 --20.49  &	 --21.79  &	 --22.51  &	 --22.91  &	 --22.92  &	 --23.20  &	 --23.19  &	 --23.10  &	 --22.62  &      38.32\\
0.1	 &2	 &	 --17.78  &	 --17.85  &	 --18.66  &	 --19.33  &	 --20.01  &	 --16.84  &	 --18.24  &	 --18.93  &	 --19.45  &	 --19.73  &	 --21.14  &	 --22.01  &	 --22.64  &	 --22.66  &	 --23.50  &	 --23.69  &	 --24.45  &	 --27.17  &      38.32\\
0.1	 &3	 &	 --18.79  &	 --18.79  &	 --19.29  &	 --19.62  &	 --20.04  &	 --18.37  &	 --19.01  &	 --19.33  &	 --19.54  &	 --19.67  &	 --20.72  &	 --21.32  &	 --21.75  &	 --21.76  &	 --22.35  &	 --22.29  &	 --23.10  &	 --26.54  &      38.32\\
0.1	 &4	 &	 --19.44  &	 --19.49  &	 --20.04  &	 --20.45  &	 --20.80  &	 --18.25  &	 --19.81  &	 --20.17  &	 --20.33  &	 --20.36  &	 --21.20  &	 --21.64  &	 --21.96  &	 --21.98  &	 --22.46  &	 --22.52  &	 --23.08  &	 --23.88  &      38.32\\
0.2	 &1	 &	 --17.02  &	 --17.29  &	 --18.99  &	 --19.85  &	 --20.61  &	 --15.55  &	 --17.93  &	 --19.46  &	 --19.97  &	 --20.41  &	 --21.82  &	 --22.50  &	 --23.13  &	 --23.12  &	 --23.39  &	 --23.51  &	 --23.37  &	 --23.05  &      39.96\\
0.2	 &2	 &	 --17.37  &	 --17.49  &	 --18.57  &	 --19.20  &	 --19.97  &	 --16.75  &	 --17.97  &	 --18.80  &	 --19.35  &	 --19.69  &	 --21.16  &	 --21.96  &	 --22.79  &	 --22.80  &	 --23.48  &	 --23.88  &	 --24.28  &	 --26.85  &      39.96\\
0.2	 &3	 &	 --18.68  &	 --18.66  &	 --19.24  &	 --19.67  &	 --20.09  &	 --18.37  &	 --18.93  &	 --19.39  &	 --19.56  &	 --19.70  &	 --20.81  &	 --21.36  &	 --21.93  &	 --21.94  &	 --22.36  &	 --22.61  &	 --22.85  &	 --26.10  &      39.96\\
0.2	 &4	 &	 --18.97  &	 --19.10  &	 --20.07  &	 --20.44  &	 --20.89  &	 --17.85  &	 --19.62  &	 --20.15  &	 --20.39  &	 --20.46  &	 --21.32  &	 --21.74  &	 --22.17  &	 --22.18  &	 --22.62  &	 --22.73  &	 --23.08  &	 --23.98  &      39.96\\
0.3	 &1	 &	 --16.61  &	 --16.80  &	 --18.56  &	 --19.69  &	 --20.51  &	 --14.61  &	 --17.38  &	 --19.25  &	 --19.88  &	 --20.31  &	 --21.79  &	 --22.55  &	 --23.27  &	 --23.28  &	 --23.54  &	 --23.71  &	 --23.65  &	 --23.46  &      40.96\\
0.3	 &2	 &	 --17.08  &	 --17.16  &	 --18.43  &	 --19.10  &	 --19.88  &	 --16.73  &	 --17.64  &	 --18.73  &	 --19.20  &	 --19.68  &	 --21.12  &	 --21.96  &	 --22.85  &	 --22.89  &	 --23.52  &	 --24.05  &	 --24.32  &	 --26.36  &      40.96\\
0.3	 &3	 &	 --18.55  &	 --18.56  &	 --19.20  &	 --19.67  &	 --20.10  &	 --18.40  &	 --18.85  &	 --19.41  &	 --19.54  &	 --19.76  &	 --20.84  &	 --21.43  &	 --22.03  &	 --22.05  &	 --22.45  &	 --22.87  &	 --22.87  &	 --25.39  &      40.96\\
0.3	 &4	 &	 --18.46  &	 --18.68  &	 --20.07  &	 --20.45  &	 --20.93  &	 --17.48  &	 --19.29  &	 --20.16  &	 --20.39  &	 --20.55  &	 --21.40  &	 --21.87  &	 --22.33  &	 --22.33  &	 --22.69  &	 --22.94  &	 --23.14  &	 --24.05  &      40.96\\
\enddata

\tablecomments{\small{The electronic table supplies the absolute
magnitude of the four template model as a function of redshift, along
with the distance modulus DM. A complete version of this tables can be
found on the electronic edition of the journal. See the caption of
Table \ref{tab:3s_modmag} for directions on how use the tables to
determine photometric redshifts and $K$ corrections.}}
\end{deluxetable}

%% file: tab5.tex
\begin{deluxetable}{l c c c c c c}

\tablehead{Templates and Sample & $\sigma_z/(1+z)$ & $\Delta z$ & 68.3\% & 95.5\% & 99.7\% & Median}

\tablecaption{Photometric - Spectroscopic Redshift Comparison Summary\label{tab:zphot_sum}}
\tabletypesize{\small}
\tablewidth{0pt}
\tablecolumns{7}

\startdata
3 template/complete sample   & 0.060 & 0.038 & 0.039 & 0.126 & 0.348 & 0.016\\
3 template/$\chi^2$ limited  & 0.044 & 0.030 & 0.033 & 0.088 & 0.245 & 0.020\\
4 template/complete sample   & 0.060 & 0.039 & 0.042 & 0.119 & 0.335 & 0.014\\
4 template/$\chi^2$ limited  & 0.048 & 0.033 & 0.037 & 0.092 & 0.268 & 0.023\\
\enddata

\tablecomments{{Summary of the photometric redshifts calculations for
  the AGES photometric galaxy sample. For each case discussed in
  \S~\ref{ssec:photoz_res} we present $\sigma_z/(1+z)$ (as defined in
  eq. [\ref{eq:sigmaz}]), $\Delta z$ (the 95\% clipped distribution
  $\sigma_z/(1+z)$), the ranges of $|z_p-z_s|/(1+z_s)$ encompassing 68.3,
  95.5 and 99.7\% of the distribution and the median value of
  $z_p-z_s$.}}
\end{deluxetable}

%% file: tab6.tex
\begin{deluxetable}{c c c r c}

  \tablehead{$\hat{e}$ range &$z$ range & $M^{*}\ -\ 5\log h$ & \colhead{$\alpha$} & 
$\phi^{*} (h^3\ \rm Mpc^{-3}\ \rm mag^{-1}$)}

  \tablecaption{NDWFS Bo\"otes Field B-band Luminosity Functions\label{tab:schech_fits}}
  \tabletypesize{\footnotesize}
  \tablewidth{0pt}
  \tablecolumns{7}

  \startdata
  $0.8 < \hat{e} < 1.0$ &   $0.0 < z < 0.2$ &  $ -18.99\ \pm\ 0.12$ &  $   0.22\ \pm\ 0.16$ &  $   8.95\ \pm\ 0.38\ \times 10^{-3}$\\
                        &   $0.2 < z < 0.4$ &  $ -19.48\ \pm\ 0.07$ &  $   0.21\ \pm\ 0.10$ &  $   4.35\ \pm\ 0.11\ \times 10^{-3}$\\
                        &   $0.4 < z < 0.6$ &  $ -19.68\ \pm\ 0.02$ &  $   0.21\ \pm\ 0.00$ &  $   3.86\ \pm\ 0.07\ \times 10^{-3}$\\
                        &   $0.6 < z < 0.8$ &  $ -19.73\ \pm\ 0.03$ &  $   0.21\ \pm\ 0.00$ &  $   2.15\ \pm\ 0.08\ \times 10^{-3}$\\
  $0.4 < \hat{e} < 0.8$ &   $0.0 < z < 0.2$ &  $ -19.00\ \pm\ 0.17$ &  $  -0.64\ \pm\ 0.19$ &  $   8.69\ \pm\ 1.28\ \times 10^{-3}$\\
                        &   $0.2 < z < 0.4$ &  $ -19.45\ \pm\ 0.08$ &  $  -0.23\ \pm\ 0.10$ &  $   5.19\ \pm\ 0.22\ \times 10^{-3}$\\
                        &   $0.4 < z < 0.6$ &  $ -19.71\ \pm\ 0.02$ &  $  -0.23\ \pm\ 0.00$ &  $   5.33\ \pm\ 0.10\ \times 10^{-3}$\\
                        &   $0.6 < z < 0.8$ &  $ -19.73\ \pm\ 0.02$ &  $  -0.23\ \pm\ 0.00$ &  $   4.64\ \pm\ 0.16\ \times 10^{-3}$\\
  $0.0 < \hat{e} < 0.4$ &   $0.0 < z < 0.2$ &  $ -18.86\ \pm\ 0.07$ &  $  -1.30\ \pm\ 0.02$ &  $  13.85\ \pm\ 1.07\ \times 10^{-3}$\\
                        &   $0.2 < z < 0.4$ &  $ -19.22\ \pm\ 0.08$ &  $  -0.67\ \pm\ 0.11$ &  $  11.00\ \pm\ 0.71\ \times 10^{-3}$\\
                        &   $0.4 < z < 0.6$ &  $ -19.72\ \pm\ 0.02$ &  $  -0.67\ \pm\ 0.00$ &  $   9.14\ \pm\ 0.19\ \times 10^{-3}$\\
                        &   $0.6 < z < 0.8$ &  $ -19.79\ \pm\ 0.02$ &  $  -0.67\ \pm\ 0.00$ &  $   9.26\ \pm\ 0.34\ \times 10^{-3}$\\
  $0.0 < \hat{e} < 1.0$ &   $0.0 < z < 0.2$ &  $ -19.64\ \pm\ 0.05$ &  $  -1.23\ \pm\ 0.02$ &  $  15.59\ \pm\ 0.86\ \times 10^{-3}$\\
                        &   $0.2 < z < 0.4$ &  $ -19.55\ \pm\ 0.05$ &  $  -0.54\ \pm\ 0.06$ &  $  18.07\ \pm\ 0.67\ \times 10^{-3}$\\
                        &   $0.4 < z < 0.6$ &  $ -19.87\ \pm\ 0.01$ &  $  -0.54\ \pm\ 0.00$ &  $  17.13\ \pm\ 0.21\ \times 10^{-3}$\\
                        &   $0.6 < z < 0.8$ &  $ -20.00\ \pm\ 0.01$ &  $  -0.54\ \pm\ 0.00$ &  $  12.10\ \pm\ 0.25\ \times 10^{-3}$\\
  \enddata
  \tablecomments{\small{Best fit Schechter function parameters for the
  luminosity functions of the NDWFS Bo\"otes field.}}

\end{deluxetable}

%% file: ms.bbl
\begin{thebibliography}{99}

\bibitem[Becker et al., 1995]{first}
Becker, R.H., White R.L. \& Helfand, D.J., 1995, \aj, 450, 559

\bibitem[Beckwith et al., 2006]{udf}
Beckwith, S.V.W. et al.\ 2006, \aj, 132, 1729

\bibitem[Bell et al., 2004]{bell04}
Bell, E.F. et al.\ 2004, \apj, 608, 752

\bibitem[Ben\'itez, 2000]{benitez00}
Ben\'itez, N.\ 2000, \apj, 563, 571

\bibitem[{Bertin \& Arnouts}, 1996]{sextractor96}	
Bertin, E. \& Arnouts, S.\ 1996, \aaps, 117, 393

\bibitem[Bessell, 1990] {bessell90}
Bessel, M.S.\ 1990, \pasp, 102, 1181

\bibitem[Blanton et al., 2003a]{blanton03}
Blanton, M.R., Brinkmann, J., Csabai, I., Doi, M., Eisenstein, D.,
Fukugita, M., Gunn, J., Hogg, D. \& Schlegel, D.\ 2003a, \aj, 125, 2348

\bibitem[Blanton et al., 2003b]{blanton03b}
Blanton, M.R. et al.\ 2003b, \apj, 594, 186

\bibitem[Blanton et al., 2006]{blanton06}
Blanton, M.R. \& Roweis, S.\ 2006, submitted (astro-ph/0606170)

\bibitem[Bolzonella et al., 2000]{bolzonella00}
Bolzonella, M., Miralles, J.-M. \& Pelló, R.\ 2000, \aap, 363, 476

\bibitem[Brodwin et al., 2006]{brodwin06}
Brodwin, M. et al.\ 2006, \apj, in press (astro-ph/0607450)

\bibitem[Brown et al., 2007]{brown07}
Brown, M.J.I. et al.\ 2007, \apj, 654, 858

\bibitem[Brown et al., in prep.]{browninprep}
Brown, M. et al.\ in preparation. 

\bibitem[Brunner et al., 1999]{brunner99}
Brunner, R.J, Connolly, A.J. \& Szalay, A.S.\ 1999, \apj, 516, 563

\bibitem[{Bruzual \& Charlot}, 1993]{bc93}
Bruzual,G. Charlot, Stephane 1993ApJ...405..538B

\bibitem[{Bruzual \& Charlot}, 2003]{bc03}
Bruzual, G. \& Charlot, S.\ 2003, \mnras, 344, 1000

\bibitem[Budavari et al., 2000]{budava00} 
Budavari T., Szalay, A.S., Connolly, A.J., Csabai, I. and Dickinson,
M.\ 2000, \aj, 120, 1588

\bibitem[{Coleman, Wu \& Weedman}, 1980]{cww80}
Coleman, G.D., Wu, C.-C. \& Weedman, D.W.\ 1980, \apjs, 43, 393 

\bibitem[{Collister \& Lahav}, 2004]{annz}
Collister, A.A. \& Lahav, O.\ 2004, \pasp, 116, 345

\bibitem[Cool, 2006]{cool06}
Cool, R.J.\ 2006, \apjs, in press (astro-ph/0611508)

\bibitem[Condon et al., 1998]{nvss}
Condon, J.J. et al.\ 1998, AJ, 115, 1693

\bibitem[Connolly et al., 1995]{connolly95}
Connolly, A.J. et al.\ 1995, \aj, 110, 2655

\bibitem[Csabai et al., 2000]{csabai00}
Csabai, I., Connolly, A.J., Szalay, A.S. and Budavari, T.\ 2000, \aj,
119, 69

\bibitem[Csabai et al., 2003]{csabai03}
Csabai, I., et al.\ 2003, \aj, 125, 580

\bibitem[{Devriendt, Guiderdoni \& Sadat}, 1999]{devriendt99}
Devriendt, J.E.G., Guiderdoni, B. \& Sadat, R.\ 1999, \aap, 350, 381

\bibitem[Dey et al., 2005]{dey05}
Dey, A. et al.\ 2005, submitted.

\bibitem[de Vries et al., 2002]{devries02}
de Vries, W.H. et al.\ 2002, \aj, 123, 1784

\bibitem[Dickinson et al., 2003]{goods} 
Dickinson, M., Giavalisco, M., and the GOODS Team\ 2003, {\it{The Great
Observatories Origins Deep Survey}}, in ``The Mass of Galaxies at Low and
High Redshift'' Proceedings of the ESO Workshop held in Venice, Italy,
24-26 October 2001; eds. R. Bender \& A. Renzini, p. 324

\bibitem[Donas et al., 1995]{donas95}
Donas, J., Milliard, B. \& Laget, M.\ 1995, \aap, 303, 661

\bibitem[Elston et al., 2006]{flamex06}
Elston, R.J., Gonzalez, A. H. et al.\ 2006, \apj, 639, 816

\bibitem[Eisenhardt et al., 2004]{irac04}
Eisenhardt, P.R. et al.\ 2004, \apjs, 154, 48

\bibitem[Eisenstein et al., in prep.]{eisenstein07}
Eisenstein, D.J. et al.\ 2007, in preparation

\bibitem[Fabricant et al., 2005]{fabricant05}
Fabricant, D. et al.\ 2005, \pasp, 117, 1411

\bibitem[Fazio et al., 2004]{fazio04}
Fazio, G.G., et al.\ 2004, \apjs, 154, 10

\bibitem[Feldmann et al., 2006]{zebra06} 
Feldmann, R. et al.\ 2006, \mnras, in press (astro-ph/0609044)

\bibitem[{Fioc \& Rocca-Volmerange}, 1997]{pegase97}
Fioc, M. \& Rocca-Volmerange, B.\ 1997, \aap, 326, 950

\bibitem[Hogg, 1999]{hogg99}
Hogg, D.\ 1999, astro-ph/9905116

\bibitem[Hogg et al., 2002]{hogg02}
Hogg, D., Baldry, I.K., Blanton, M.R. \& Eisenstein, D.J., 2002,
astro-ph/0210394

\bibitem[Jannuzi \& Dey, 1999]{ndwfs99} 
Jannuzi, B. T. \& Dey, A.\ 1999, ASP Conference Series, Vol. 191,
p. 111

\bibitem[Jannuzi et al., 2005]{jannuzi05}
Jannuzi, B.T. et al.\ 2005, submitted.

\bibitem[Kauffmann et al., 2003]{kauffmann03}
Kauffmann, G. et al.\ 2003, MNRAS, 341, 33

\bibitem[Kinney et al., 1996]{kinney96}
Kinney, A.L. et al.\ 1996, \apj, 1996, 467, 38

\bibitem[Kochanek et al., in prep]{ages}
Kochanek, C.S. et al. in preparation.

\bibitem[{Lawson \& Hanson}, 1974]{lawson74} 
Lawson, C.L. \& Hanson, R.J.\ 1974, \textit{Solving Least Squares
Problems}, PrenticeHall.

\bibitem[Lee et al., 2006]{lee06}
Lee, K. et al., 2006, \apj, 642,63

\bibitem[Lin et al., 1996]{lcrslf}
Lin, H., Kirshner, R.P., Shectman, S.A., Landy, S.D., Oemler, A.,
Tucker, D.L. \& Schechter, P. L.\ 1996, \apj, 464, 60

\bibitem[Madgwick et al., 2003]{madgwick03}
Madgwick, D.S. et al.\ 2003, \apj, 599, 997

\bibitem[Martin et al., 2005]{martin05}
Martin, D.C. et al.\ 2005, \apj, 619L, 1

\bibitem[Murray et al., 2005]{murray05}
Murray, S.S. et al.\ 2005, \apjs, 161, 1

\bibitem[{Oke \& Sandage}, 1968]{oke68}
Oke, J.B. \& Sandage, A.\ 1968, \apj, 154, 21

\bibitem[Ouchi et al., 2005]{ouchi05}
Ouchi, M. et al.\ 2005, \apj, 635L, 117

\bibitem[{Padmanabhan \& Ray}, 2006]{padma06}
Padmanabhan, T. \& Ray, Suryadeep\ 2006, \mnras, 372, 53

\bibitem[Schechter, 1976]{schech76}
Schechter, P.\ 1976, \apj, 203, 297

\bibitem[Schlegel et al., 1998]{schlegel98}
Schlegel, D.J., Finkbeiner, D.P. \& Davis, M.\ 1998, \apj, 500, 525

\bibitem[Schmidt, 1968]{schmidt68}
Schmidt, M.\ 1968, \apj, 151, 393

\bibitem[Stern et al., 2005]{stern05}
Stern, D. et al.\ 2005, \apj, 631, 163

\bibitem[Reach et al., 2005]{reach05}
Reach, W.T. et al.\ 2005, \pasp, 117, 978

\bibitem[Rengelink et al., 1997]{wenss}
Rengelink, R.B. et al.\ 1997, \aap, 124, 259

\bibitem[Scoville et al., 2006]{cosmos06}
Scoville N. et al.\ 2006, \apjs, COSMOS Special Issue

\bibitem[Skrutskie et al., 2006]{2mass}
Skrutskie, M.F. et al.\ 2006,\aj, 131, 1163

\bibitem[Strateva et al., 2001]{strateva01}
Strateva, I. et al.\ 2001, \aj, 122, 186

\bibitem[Wang et al., 1998]{wang98}
Wang, Y., Bahcall, N. \& Turner, E.L.\ 1998, \aj, 116, 2081

\bibitem[Weedman et al., 2006]{weedman06}
Weedman, D.W. et al.\ 2006, \apj, 651, 101

\bibitem[Weiner et al., 2005]{weiner05}
Weiner, B.J. et al.\ 2005, \apj, 620, 595

\bibitem[White et al., 2007]{white07}
White, M. et al.\ 2007, \apj, 655L, 69

\bibitem[Wolf et al., 2003]{wolf03} 
Wolf, C., Meisenheimer, K., Rix, H.-W., Roch, A., Dye, S., and
Kleinheinrich, M.\ 2003, \aap, 401, 73

\bibitem[York et al., 2000]{sdss}
York D. et al.\ 2000, \aj, 120, 1579

\bibitem[Zehavi et al., 2005]{zehavi05}
Zehavi, I. et al.\ 2005, \apj, 630, 1

\end{thebibliography}
